\documentclass[useAMS,usenatbib,usegraphicx]{mn2e}
\usepackage{color}
\usepackage{psfig}
\usepackage{epsfig}
\usepackage{amsbsy}  
\usepackage{amssymb}
\usepackage{amsmath}

\definecolor{ndcol}{rgb}{0.75,0.0,0.0}

\usepackage[colorlinks,citecolor=black,linkcolor=black]{hyperref}

\newcommand*{\unit}[1]{\ensuremath{\mathrm{\, #1}}}
\newcommand*{\erg}{\unit{erg}}
\newcommand*{\keV}{\unit{keV}}
\newcommand*{\second}{\unit{s}}

\newcommand*{\cm}{\unit{cm}}

\newcommand*{\Mpc}{\unit{Mpc}}

\newcommand{\Hunit}{~{\rm km}\allowbreak{}~{\rm s}^{-1}\allowbreak{} {\rm Mpc}^{-1} }

\newcommand*{\E}[1]{\ensuremath{\times 10^{#1}}}
\newcommand*{\logTen}{\ensuremath{\log_{10}}}
\newcommand*{\expectation}[1]{\ensuremath{\left\langle #1 \right\rangle}}

\newcommand*{\Msun}{\ensuremath{\, M_{\odot}}}
\newcommand*{\mysub}[2]{\ensuremath{#1_{\mathrm{#2}}}}

\defcitealias{Rapetti:10}{R10}
\defcitealias{Rapetti:09}{R09}
\defcitealias{Mantz:08}{M08}
\defcitealias{Allen:08}{A08}

\newcommand*{\Chandra}{{\it{Chandra}}}

\title[Growth and expansion from clusters, the CMB and
galaxies] {A combined measurement of cosmic growth and expansion from
  clusters of galaxies, the CMB and galaxy clustering}

\author[D.~Rapetti et al.]{David~Rapetti${}^{1}$\thanks{Email: drapetti@dark-cosmology.dk}, Chris~Blake${}^{2}$, Steven~W.~Allen${}^{3,4}$, Adam~Mantz${}^{5}$, David~Parkinson${}^{6}$ \newauthor and Florian~Beutler${}^{7,8}$\vspace{0.05in}\\
  $^1$ Dark Cosmology Centre, Niels Bohr Institute, University of Copenhagen, Juliane Maries Vej 30, 2100 Copenhagen, Denmark\\
  $^2$ Centre for Astrophysics \& Supercomputing, Swinburne University of Technology, P.O. Box 218, Hawthorn, VIC 3122, Australia\\
  $^3$ Kavli Institute for Particle Astrophysics and Cosmology, Stanford University, 452 Lomita Mall, Stanford, CA 94305-4085, USA \\
  $^4$ SLAC National Accelerator Laboratory, 2575 Sand Hill Road, Menlo Park 94025, CA, USA\\
  $^5$Kavli Institute for Cosmological Physics, University of Chicago, 5640 South Ellis Avenue, Chicago, IL 60637, USA\\
  $^6$School of Mathematics and Physics, University of Queensland, Brisbane, QLD 4072, Australia\\
  $^7$International Centre for Radio Astronomy Research, University of Western Australia, 35 Stirling Highway, Perth WA 6009, Australia\\
  $^8$Lawrence Berkeley National Laboratory, 1 Cyclotron Road, Berkeley, CA 94720, USA}

\begin{document}
\date{Accepted 2013 March 21. Received 2013 March 21; in original form 2012 May 18}
\pagerange{\pageref{firstpage}--\pageref{lastpage}} \pubyear{2011}

\maketitle
\label{firstpage}

\begin{abstract}
  Combining galaxy cluster data from the {\it ROSAT} All-Sky Survey
  and the \Chandra{} X-ray Observatory, cosmic microwave background
  data from the {\it Wilkinson Microwave Anisotropy Probe}, and galaxy
  clustering data from the WiggleZ Dark Energy Survey, the 6-degree
  Field Galaxy Survey and the Sloan Digital Sky Survey III, we test
  for consistency the cosmic growth of structure predicted by General
  Relativity (GR) and the cosmic expansion history predicted by the
  cosmological constant plus cold dark matter paradigm
  ($\Lambda$CDM). The combination of these three independent, well
  studied measurements of the evolution of the mean energy density and
  its fluctuations is able to break strong degeneracies between model
  parameters. We model the key properties of cosmic growth with the
  normalization of the matter power spectrum, $\sigma_8$, and the
  cosmic growth index, $\gamma$, and those of cosmic expansion with
  the mean matter density, $\Omega_{\rm m}$, the Hubble constant,
  $H_0$, and a kinematical parameter equivalent to that for the dark
  energy equation of state, $w$. For a spatially flat geometry,
  $w=-1$, and allowing for systematic uncertainties, we obtain
  $\sigma_8=0.785\pm0.019$ and $\gamma=0.570^{+0.064}_{-0.063}$ (at
  the $68.3$ per cent confidence level). Allowing both $w$ and
  $\gamma$ to vary we find $w=-0.950^{+0.069}_{-0.070}$ and
  $\gamma=0.533\pm0.080$. To further tighten the constraints on the
  expansion parameters, we also include supernova, Cepheid variable
  and baryon acoustic oscillation data. For $w=-1$, we have
  $\gamma=0.616\pm0.061$. For our most general model with a free $w$,
  we measure $\Omega_{\rm m}=0.278^{+0.012}_{-0.011}$,
  $H_0=70.0\pm1.3\Hunit$ and $w=-0.987^{+0.054}_{-0.053}$ for the
  expansion parameters, and $\sigma_8=0.789\pm0.019$ and
  $\gamma=0.604\pm0.078$ for the growth parameters. These results are
  in excellent agreement with GR+$\Lambda$CDM ($\gamma\simeq 0.55$;
  $w=-1$) and represent the tightest and most robust simultaneous
  constraint on cosmic growth and expansion to date.
\end{abstract}


\section{Introduction}
\label{sec:introduction}

The unexpected measurement from type Ia supernova (SNIa) data of
late-time cosmic acceleration by \cite{Riess:98} and
\cite{Perlmutter:99} initiated a series of theoretical and
observational efforts to unveil the nature of its underlying
cause. However, to this day it is still unclear whether the origin of
this phenomenon is due to a new energy component, spurious
cosmological assumptions, or modifications of gravity at large
scales. A number of theoretical approaches and observational probes
have been developed to investigate these different possibilities
\citep[for recent reviews see][]{Copeland:06, Frieman:08, Allen:11,
  Clifton:11, Weinberg:12}. Current data on the energy content,
geometry, and expansion and growth histories of the Universe do not
show any deviation from the standard cosmological paradigm,
$\Lambda$CDM \citep{Allen:08, Mantz:08, Mantz:10a, Vikhlinin:09,
  Percival:09, Reid:12, Conley:11, Blake:11a, Blake:11b, Blake:12,
  Suzuki:12, Hinshaw:12}. However, the cosmological constant model
suffers from well-known, serious theoretical problems that present-day
dark energy models have not been able to improve upon. For modified
gravity models, various approaches have been
developed\footnote{Similar approaches can also be used to study
  clustering dark energy models.}: parameterized frameworks
\citep{Hu:07, Bertschinger:08, Amin:08}, consistency tests of GR
\citep{Linder:05, Linder:07, DiPorto:07, Zhang:07, Nesseris:08,
  Acquaviva:08}, and alternative theories of gravity \citep{Dvali:00,
  Carroll:03b, Arkani-Hamed:04, Nicolis:09, deRham:11}. Recent works
have used a variety of experiments and data sets to constrain gravity
properties and models and found no significant deviations from GR
\citep[see e.g.][]{Schmidt:09d, Rapetti:09, Rapetti:10, Reyes:10,
  Daniel:10, Zhao:10, Zhao:11, Giannantonio:10, Wojtak:11, Hojjati:11,
  Lombriser:11, Hudson:12, Basilakos:12, Samushia:12, Simpson:12}. To
further test the overall standard paradigm, GR+$\Lambda$CDM, it is
crucial to use data sets able to robustly constrain the key properties
of the model, and to combine complementary data sets to break the
degeneracies between the model parameters.

\citet[][hereafter R10]{Rapetti:10} tested GR using {\it ROSAT}
All-Sky Survey (RASS) and \Chandra{} X-ray Observatory (CXO) data of
cluster abundance and scaling relations \cite[][hereafter
M10a,b]{Mantz:10a, Mantz:10b}. R10 obtained strong constraints on GR,
even when marginalizing over conservative systematic and astrophysical
modeling uncertainties in the evolution of the cluster X-ray
luminosity-mass relation. When combining the cluster growth data with
measurements of the anisotropies in the cosmic microwave background
(CMB) from the {\it Wilkinson Microwave Anisotropy Probe} \citep[{\it
  WMAP};][and companion papers]{Spergel:03, Spergel:07, Komatsu:09,
  Dunkley:09, Komatsu:11, Hinshaw:12, Bennett:12}, they highlighted a
large degeneracy between $\gamma$ and $\sigma_8$, which limited the
constraints on each parameter individually. Here we include
complementary data that break this degeneracy. In particular, we use
measurements of the growth rate and the Hubble parameter from joint
redshift space distortions (RSD) and Alcock-Paczynski (AP) effect
constraints \cite[][hereafter B11]{Blake:11c} from the WiggleZ Dark
Energy Survey \cite[WiggleZ;][]{Drinkwater:10}. We also use a low
redshift RSD constraint \citep{Beutler:12} from the final data release
(DR3) of the 6-degree Field Galaxy Survey \cite[6dFGS;][]{Jones:09}
and an RSD and AP constraint from the latest data release (DR9) of the
Sloan Digital Sky Survey (SDSS) III Baryon Oscillation Spectroscopic
Survey \cite[BOSS;][]{Reid:12}.

In addition to our primary data sets, and to tighten the constraints
on the expansion parameters, we also present results including data
from the Union II SNIa sample \citep{Suzuki:12}, baryon acoustic
oscillation (BAO) measurements from a combined analysis
\citep{Percival:09} of 2-degree Field Galaxy Redshift Survey
\cite[2dFGRS;][]{Colless:03} and SDSS-II DR7 \citep{Abazajian:09} data
as well as from a recent analysis \citep{Reid:12} of SDSS-III BOSS
data at a higher redshift, and $H_0$ measurements from the Supernovae
and $H_0$ for the Equation of State program \cite[SH0ES;][]{Riess:11}.

We find that by combining cluster, CMB and galaxy data we are able to
break the key degeneracies between $\gamma$ and $\sigma_8$ and obtain
tight and robust constraints on cosmic expansion and growth. We model
the expansion primarily with $\Omega_{\rm m}$, $H_0$, and $w$ and the
growth with $\sigma_8$ and $\gamma$. We find that, individually, the
CMB and galaxy data have large degeneracies in the growth plane but
that, crucially, these degeneracies are nearly orthogonal. The
individual and combined constraints from cluster, CMB and galaxy data
are consistent with one another, making this a very robust
measurement, and in good agreement with GR and $\Lambda$CDM. While
individually clusters provide the tightest constraints in the growth
plane, the combination of clusters, the CMB and galaxies provides
significantly improved constraints and arguably the most robust
measurement of cosmic structure growth to date.

\section{Cosmological model}
\label{sec:cosmo}

We adopt a purely phenomenological model to conveniently test the
consistency of current observations with both the cosmic expansion
history and the cosmic growth history predicted by $\Lambda$CDM+GR.

Our model assumes neither the existence of a new component, dark
energy, nor a modification of GR. Instead, the parameters of the model
represent departures from key kinematical and dynamical features of
$\Lambda$CDM+GR. Deviations from such benchmarks would indicate
disagreement of the observed evolution of the background and density
perturbations with the standard cosmological paradigm.\footnote{Note
  that different physical scenarios can cause similar departures from
  this paradigm. For example, specific models of clustering or
  interacting dark energy and of modified gravity might provide
  similar deviations from the density perturbations of
  $\Lambda$CDM+GR.}

\subsection{Cosmic expansion history}
\label{sec:cosmo:exp}

We model the expansion history using the evolution parameter
$E(a)\equiv H(a)/H_0$, where $H(a)$ is the Hubble parameter as a
function of the scale factor $a$ and $H_0$ its present-day value. We
parameterize $E(a)$ as follows
 
\begin{equation}
  E(a)=\left[\Omega_{\rm m} \,a^{-3}+(1-\Omega_{\rm m}) \,a^{-3(1+w)}\right]^{1/2}\,.
\label{eq:Ea}
\end{equation}

\noindent $\Omega_{\rm m}$ is the present, mean matter density in
units of the critical density of the Universe and $w$ a kinematical
parameter inspired by the dark energy equation of state. Since for our
test we do not assume any particular scenario for cosmic acceleration,
such as dark energy, we use $w$ only to conveniently fit expansion
history data, matching $\Lambda$CDM for $w=-1$. Below, we present
results for two expansion models, $w=-1$ ($\Lambda$CDM) and $w$
constant ($w$CDM). For both cases, we assume a spatially flat geometry
(i.e., the curvature energy density $\Omega_{\rm k}=0$)\footnote{Using
  cluster, CMB and SNIa data, \cite{Rapetti:09} found a negligible
  correlation between $\Omega_{\rm k}$ and $\gamma$. They also showed
  that the constraints on $\gamma$ were not significantly weaker when
  including $\Omega_{\rm k}$ as a free parameter. Note also that if
  $\Omega_{\rm k}$ were included as a free parameter, an extension of
  equation~\ref{eq:fa} proposed by \cite{Gong:09} would fit better the
  predictions from GR.}.

\subsection{Cosmic growth and cluster abundance}
\label{sec:cosmo:growth}

We model the growth history at late times by parameterizing the linear
growth rate of density perturbations on large scales, $f(a)$, as a
power law of the evolving mean matter density, $\Omega_{\rm
  m}(a)=\Omega_{\rm m} a^{-3} E(a)^{-2}$, such as \citep{Peebles:80,
  Wang:98, Linder:05}

\begin{equation}
  f(a)\equiv \frac{d\ln\delta}{d\ln a}=\Omega_{\rm m} (a)^{\gamma}\,,
\label{eq:fa}
\end{equation}

\noindent where $\gamma$ is the growth index\footnote{Many models of
  modified gravity predict a growth index that varies with time and
  length scale, $\gamma(a,k)$. Note again, though, that here we do not
  use this parameter as a diagnostic of the true theory of gravity,
  but rather as a consistency test for GR.}, for which we recover GR
when $\gamma\simeq 0.55$.\footnote{For current results, this value is
  a good approximation to be used as a GR reference. At higher
  accuracy, though, the growth index of GR has relatively small
  redshift and background parameter dependencies \citep[see
  e.g.][]{Polarski:08}.}  $\delta\equiv \delta\rho_{\rm m}/\rho_{\rm
  m}$ is the ratio of the comoving matter density fluctuations,
$\delta\rho_{\rm m}$, with respect to the cosmic mean, $\rho_{\rm
  m}$. While at early times we assume GR, for $z<z_{\rm t}$ we obtain
$\delta(z)$ from equation~\ref{eq:fa} using as an initial condition
$\delta(z_{\rm t})$ calculated within GR. Normalizing $\delta(z)$ to
$\delta(z_{\rm t})$, we obtain the growth factor,
$D(z)\equiv\delta(z)/\delta(z_{\rm t})$. Here we use $z_{\rm t}=30$,
which is well within the dark matter dominated era, when $f(a)\sim 1$
for both the $\gamma$-model (equation~\ref{eq:fa}) and GR. We then
calculate the matter power spectrum of such fluctuations for a given
wavenumber, $k$, as

\begin{equation}
  P(k,z)\propto k^{n_{\rm s}} T^2(k,z_{\rm t}) D(z)^2\,,
\label{eq:powers}
\end{equation}

\noindent where $T(k,z_{\rm t})$ is the matter transfer function of GR
in the synchronous gauge at redshift $z_{\rm t}$ and $n_{\rm s}$ the
primordial scalar spectral index. 

The variance of the linearly evolved density field, smoothed by a
spherical top-hat window function of comoving radius $R$ enclosing
mass $M=4\pi\rho_{\rm m}R^3/3$, is

\begin{equation}
  \sigma^2(M,z) = \frac{1}{2\pi^2} \int_0^\infty k^2 P(k,z) |W_{\rm M}(k)|^2 dk\,,
\label{eq:var}
\end{equation}

\noindent where $W_{\rm M}(k)$ is the Fourier transform of the window
function. From this expression, $\sigma_8^2$ is defined as the $z=0$
variance in the density field at scales of $8h^{-1}\Mpc$, where
$\sigma_8$ is widely used as a parameter for the normalization of the
matter power spectrum.

Here we use $\sigma(M,z)$ to calculate the abundance of dark matter
halos as a function of mass and redshift

\begin{equation}
  n(M,z)=\int_{0}^{M}\mathcal{F}(\sigma,z)\,\frac{\rho_{\rm m}}{M'}\,\frac{d\ln\sigma^{-1}}{dM'}\,dM'\,,
\label{eq:nmz}
\end{equation}

\noindent where $\mathcal{F}(\sigma,z)$ is a convenient fitting formula
obtained from large N-body simulations of dark matter particles
\citep{Tinker:08},

\begin{equation}
  \mathcal{F}(\sigma,z) = A\left[\left(\frac{\sigma}{b}\right)^{-a}+1\right]e^{-c/\sigma^2}\,.
\label{eq:mf}
\end{equation}

\noindent The parameters of this formula have a generic redshift
dependence of the form $x(z)=x_0(1+z)^{\varepsilon \alpha_{\rm x}}$,
with $x$ representing $A,a,b$ or $c$. The values for each $x_0$ and
$\alpha_{\rm x}$ are given in \cite{Tinker:08}. As in M10a, we
introduce an additional parameter, $\varepsilon$, to account for
residual systematic uncertainties in the evolution of
$\mathcal{F}(\sigma,z)$ due to non-$\Lambda$CDM scenarios. Remarkably,
$\mathcal{F}(\sigma,z)$ encapsulates the non-linear cosmic growth
history and appears to be almost universal for a wide range of
cosmologies (see R10 for more details).

We marginalize over the uncertainties in the parameters of
$\mathcal{F}(\sigma,z)$, accounting for their covariance and for
additional systematic uncertainties due to e.g. the presence of
baryons following the method described in M10a. Note, though, that as
shown in M10a the uncertainties in $\mathcal{F}(\sigma,z)$ are
subdominant in the analysis. R10 also verified that $\varepsilon$ is
essentially uncorrelated with $\gamma$.

\subsection{Integrated Sachs-Wolfe effect}
\label{sec:cosmo:isw}

In our CMB analysis we include the constraint on $\gamma$ from the
Integrated Sachs-Wolfe (ISW) effect of the CMB using the method and
assumptions described by \cite{Rapetti:09, Rapetti:10}. In brief, the
low multipoles of the CMB are sensitive to the growth of cosmic
structure due to the effect of the time-varying gravitational
potentials of large scale structures on the CMB photons crossing
them. We calculate the contribution of these photons to the
temperature anisotropy power spectrum as \citep{Weller:03}

\begin{equation}
  \Delta_{l}^{\rm ISW}(k) = 2 \int dt\, {\rm e}^{-\tau(t)}
  \phi'j_{l}\left[k(t-t_{\rm 0})\right]\,,
  \label{eq:deltaisw}
\end{equation}

\noindent where $t$ is the conformal time and $t_0$ its present-day
value, $\tau$ the optical depth to reionization, $j_{l}(x)$ the
spherical Bessel function for the multipole $l$, and $\phi'$ the
conformal time variation of the gravitational potential. Taking the
derivative of the Poisson equation with respect to $t$, we calculate
the latter quantity for the $\gamma$-model\footnote{Since here we are
  testing GR, we assume no contributions to $\phi$ from the
  anisotropic stress and energy flux of the Weyl tensor
  \citep{Challinor:98}.} as $ \phi' = 4 \pi G a^2
k^{-2}\,\mathcal{H}\,\delta\rho_{\rm m}\left[1-\Omega_{\rm
    m}(a)^{\gamma}\right]$, where $\mathcal{H}$ is the conformal
Hubble parameter. Since the ISW effect is only relevant for $z<2$, as
an initial condition to solve this equation we match $\Delta_{l}^{\rm
  ISW}(k)$ to that of GR at $z_{\rm t}=2$.\footnote{For our analysis,
  the difference from calculating $\delta(z)$ using $z_{\rm t}$ equals
  to $2$ or $30$ is negligible since both redshifts are well within
  the dark matter dominated era, when $f(a)$ tends to 1 for any
  $\gamma$.}

Note, however, that the constraining power on $\gamma$ from the ISW
effect is small compared to that of the cluster data
\citep{Rapetti:09}. For the current analysis, the primary relevance of
the CMB is its ability to tightly constrain the combination of growth
parameters $\sigma_8$ and $\gamma$ (see Section~\ref{sec:results}).

\subsection{The Alcock-Paczynski effect and redshift-space distortions}
\label{sec:cosmo:reddist}

The Alcock-Paczynski test is a geometrical means of probing the
cosmological model by a comparison of the observed tangential and
radial dimensions of objects which are assumed to be isotropic in the
correct choice of model. It can be applied to the 2-point statistics
of galaxy clustering if the redshift space distortions, the principal
additional source of anisotropy, can be successfully modelled
\citep{Ballinger:96, Matsubara:96, Matsubara:00, Seo:03,
  Simpson:10}. By equating radial and tangential physical scales, the
AP test determines the observable $F(z) = (1+z) D_{\rm A}(z) H(z)/c$,
where $D_{\rm A}(z)$ is the physical angular diameter distance and $c$
is the speed of light.

In the model fit for $F(z)$, the normalized growth rate, $f
\sigma_8(z)$, is determined simultaneously. Here $f(z)$ is again the
logarithmic rate of change of the growth factor at redshift $z$ (see
equation~\ref{eq:fa}) and $\sigma_8(z)=[ D(z)/D(0) ] \, \sigma_8$. In
B11, RSD were modelled using the fitting formulae provided by
\cite{Jennings:11} to determine the density-velocity and
velocity-velocity power spectra, marginalizing over a linear bias
factor. Tests were performed to ensure that the results were not very
sensitive to the model used for the non-linear RSD, the real-space
power spectrum, or the range of scales fitted ($k_{\rm
  max}<0.2\,h\Mpc^{-1}$, for the measurements used here).

For a low-redshift survey such as 6dFGS, the Alcock-Paczynski
distortion is negligible (since distances in $h^{-1}\Mpc$ are
approximately independent of the assumed cosmological
model)\footnote{\cite{Beutler:12} calculated the uncertainties in
  $F(z)$ and showed that they are unimportant.}. For 6dFGS, the growth
rate measurement of \cite{Beutler:12} was obtained by again assuming
the model of \cite{Jennings:11} to described non-linear RSD.

For the BOSS measurements of the RSD and AP effect, the modeling of
the matter density and velocity fields was performed following the
approach of \cite{Reid:11}. The latter uses perturbation theory to
calculate the non-linear redshift space clustering of halos in the
quasilinear regime and the halo model framework to describe the
galaxy-halo relation. This model was tested against a large set of
galaxy catalogs from N-body simulations and only fit over those scales
where the quasilinear velocity field was thought to dominate the
signal and the small-scale random velocities could be simply modeled
and marginalized over.

For all the RSD and AP effect measurements employed in the paper (see
Section~\ref{sec:gal}), the parameters used to fit the 2D galaxy power
spectrum and galaxy correlation function data have negligible
covariance with the parameters in the current analysis. Also, the
linear model as well as the non-linear corrections assumed in those
analyses lie within the GR+$\Lambda$CDM paradigm tested
here\footnote{The non-linear modeling from \cite{Jennings:11} used in
  the WiggleZ and 6dFGS analyses also encompasses a range of
  quintessence dark energy models.}.

\section{Physics of the observables}
\label{sec:physics}

In this section, we describe the physical mechanisms behind the
principle degeneracies between our most relevant growth and expansion
parameters, for each of our primary observations.

\subsection{CMB anisotropies}

From the normalization and tilt of the CMB temperature anisotropy
power spectrum, we can primarily constrain the scalar amplitude and
spectral index of primordial fluctuations; from the position of its
first peak, the mean energy density of curvature and dark energy; and
from the amplitudes of the second and third peaks, those of dark and
baryonic matter. These measurements provide strong constraints on the
content of the background energy density and its linear density
fluctuations at high redshift. For a given value of the growth index,
$\gamma$, these translate into tight constraints on the amplitude of
the matter power spectrum today, $\sigma_8$. A model with faster
perturbation growth, i.e. with a small $\gamma$, implies large
fluctuations today, i.e. large $\sigma_8$, and vice-versa. This
provides a large, negative correlation between $\sigma_8$ and $\gamma$
(see Figure~\ref{fig:datasets}). At low redshift, the ISW effect of
the CMB data (see Section~\ref{sec:cosmo:isw}) constrains $\gamma$,
which is otherwise unconstrained by this data set.

\subsection{Distribution of galaxies}

From measurements of the anisotropic clustering of galaxies, we use
constraints on the product $f(z)\,\sigma_8(z)$ and on the quantity
$F(z)$, where the latter are purely expansion history constraints,
i.e. on $\Omega_{\rm m}(z)$. Both of these constraints, from RSD and
AP effect measurements respectively, are required to measure
$\gamma=\ln f(z)/\ln\Omega_{\rm m}(z)$ from galaxy data
alone.\footnote{In the same way as for the AP effect, the addition of
  the BAO constraints on $\Omega_{\rm m}(z)$ improves significantly
  the measurement of $\gamma$ for the combination gal+BAO (see the
  right panel of Figure~\ref{fig:datasets}).} The current uncertainty
on the linear galaxy bias, $b(z)$, limits the ability to measure
$\sigma_8$ from the normalization of the galaxy power spectrum, which
scales with $\sigma_8(z)\,b(z)$, and to measure $f(z)$ using RSD
constraints on $f(z)/b(z)$, as previously commonly used. As proposed
by \cite{Song:09}, here we use instead RSD constraints on
$f(z)\,\sigma_8(z)$, which are independent of $b(z)$, and obtain a
positive correlation between $\gamma$ and $\sigma_8$ (see
Figure~\ref{fig:datasets}) for a $\Lambda$CDM expansion model and data
within a relatively low-$z$ range, where $f(z)$ increases towards
$1$. The faster the perturbations grow (small $\gamma$), the smaller
the present-day perturbation amplitude, $\sigma_8$, needs to be to
provide the same amount of anisotropy in the distribution of galaxies
at redshift $z$. At higher-$z$, where $f(z)\sim 1$ and
$f(z)\,\sigma_8(z)\sim\sigma_8(z)$, the correlation between $\gamma$
and $\sigma_8$ becomes negative (see
Section~\ref{sec:physics:cl}). Adding high-$z$ data from future
missions will then help to break the large degeneracy of the current
data between $\gamma$ and $\sigma_8$.

\subsection{Cluster abundance and masses}
\label{sec:physics:cl}

For clusters, we have direct constraints on $\sigma_8(z)$ and
$\Omega_{\rm m}(z)$ from abundance, mass calibration and gas mass
fraction data (see Sections~\ref{sec:cosmo:growth}
and~\ref{sec:cl}). $\sigma_8(z)$ measurements provide us with
constraints not only on $\sigma_8(z=0)$, from the local cluster mass
function, but also on the growth rate
$f(z)=-(1+z)d\ln\sigma_8(z)/dz$. Together, the constraints on
$\sigma_8(z)$ and $\Omega_{\rm m}(z)$ constrain $\gamma$. 

The evolution of $\sigma_8(z)=\sigma_8 e^{-g(z)}$ depends on $\gamma$,
$\Omega_{\rm m}$ and $w$ as follows

\begin{eqnarray}
  g(z)&=&\int_{0}^{z}(1+z')^{-1}\left[p(z')-1\right]^{-\gamma}p(z')^{\gamma}dz'
  \label{eq:gint1}\\
  &=&(3w\gamma)^{-1}\,\left[\lambda(z)-\lambda(0)\right]\,,
  \label{eq:gint2}
\end{eqnarray}

\noindent where $\lambda(z)=\left[p(z)-1\right]^{1-\gamma
}p(z)^{\gamma}\,_{2}F_{1}\left[1,1;1+\gamma;p(z)\right]$, $_{2}F_{1}$
is a hypergeometric function, $p(z)=p_0(1+z)^{-3w}$ and
$p_0=\Omega_{\rm m}/(\Omega_{\rm m}-1)$. In practice, a negative
degeneracy between $\sigma_8$ and $\gamma$ exists due to the limited
precision of cluster mass estimates, but it is notably smaller than
those described above (see Figure~\ref{fig:datasets}). Within the
precision of the data, indistinguishable cluster count evolution can
be produced by models with e.g. $\sigma_8$ of $0.8$ and a growth rate
consistent with GR, or with a slightly larger present-day amplitude
and faster growth (smaller $\gamma$), for which $\sigma_8(z)$
decreases with $z$ a bit more steeply.

For the $\gamma$+$w$CDM model, the dependence of $\sigma_8(z)$ on the
product $w\,\gamma$ implies a negative correlation on the $w, \gamma$
plane (see Figure~\ref{fig:wconst}). Within the precision of the data,
a fast expansion history (small $w$) can be mimicked by a slow growth
history (large $\gamma$), and vice-versa.

\section{Data analysis}
\label{sec:obs}

\subsection{Galaxy cluster data}
\label{sec:cl}

For clusters we use two experiments: growth of structure (M10a,b) and
gas mass fraction \cite[$f_{\rm gas}$;][]{Allen:08}\footnote{Note that
  the cluster growth analysis employs the $f_{\rm gas}$ analysis to
  calibrate the masses for the scaling relations of
  Section~\ref{sec:scal} using gas mass as a proxy for total mass (see
  details in M10a).}.

Following the methods developed by M10a,b for the cluster growth
analysis, we self-consistently and simultaneously combine X-ray survey
and follow-up observations to obtain the best constraints possible
while accounting fully for selection biases. We employ the survey data
to determine cluster abundances and the follow-up data to calibrate
cluster masses from three observables: luminosity, temperature and
X-ray emitting gas mass. For the survey data we employ three wide-area
cluster samples drawn from RASS: the Bright Cluster Sample in the
northern sky (BCS; $z\allowbreak{}<\allowbreak{}0.3$ and $F_{\rm
  X}(0.1-2.4\keV{})\allowbreak{}>\allowbreak{}4.4\times 10^{-12} \erg
\second^{-1} \cm^{-2}$), the {\it ROSAT}-ESO Flux Limited X-ray sample in
the southern sky (REFLEX; $z\allowbreak{}<\allowbreak{}0.3$ and
$F_{\rm X}\allowbreak{}>\allowbreak{}3.0\times 10^{-12}
\erg\second^{-1} \cm^{-2}$), and the Bright Massive Cluster Survey
with $\sim 55$ per cent sky coverage (Bright MACS;
$0.3\allowbreak{}<\allowbreak{}z\allowbreak{}<\allowbreak{}0.5$ and
$F_{\rm X}\allowbreak{}>\allowbreak{}2\times 10^{-12}\allowbreak{}\erg
\second^{-1} \cm^{-2}$). To keep systematic uncertainties to a minimum
and maintain a trivial constant scaling between X-ray gas mass and
total mass, for all three samples we impose a lower luminosity cut of
$2.5\E{44}h_{70}^{-2}\allowbreak{}\erg\second^{-1}$ ($0.1-2.4\keV{}$)
leaving a total of 78 clusters from BCS; 126 clusters from REFLEX; and
34 clusters from Bright MACS. In total we use 238 clusters. For 94 of
these clusters, we employ follow-up observations from CXO or pointed
observations from {\it ROSAT} (M10b; distributed along the same redshift
range of the survey data $0<z<0.5$) to constrain simultaneously the
luminosity-mass ($L$--$M$) and temperature-mass ($T$--$M$) relations
using the model from M10b (see a brief description in
Section~\ref{sec:scal}).

For the $f_{\rm gas}$ analysis, we use the methods and data set of
\cite{Allen:08} for 42 massive, hot ($kT>5\keV{}$), dynamically
relaxed, X-ray luminous galaxy clusters spanning the redshift range
$0.05<z<1.1$.

\subsubsection{Scaling relations model}
\label{sec:scal}

We model the $L$--$M$ scaling relation as (M10b)

\begin{equation}
  \expectation{\ell(m)} = \beta_0^{\ell m} + \beta_1^{\ell m} m + \beta_2^{\ell m}\logTen(1+z)\,,
  \label{eq:mlmt}
\end{equation}

\noindent with a log-normal intrinsic scatter at a given mass of

\begin{equation}
  \sigma_{\ell m}(z) = \sigma_{\ell m} ( 1 + \sigma_{\ell m}' z )\,,
  \label{eq:scat}
\end{equation}

\noindent where
$\ell\equiv\logTen[L_{500}E(z)^{-1}/10^{44}\erg\second^{-1}]$ and
$m\equiv\logTen[E(z)M_{500}/10^{15}\Msun]$. The subscript $500$ refers
to quantities measured within radius $r_{500}$, at which the mean,
enclosed density is 500 times the critical density of the Universe at
redshift $z$. We model the $T$--$M$ scaling relation
$\expectation{t(m)}$, where $t\equiv
\logTen\left(\mysub{kT}{500}/\keV\right)$, and its scatter
$\sigma_{tm}(z)$ using the same equations~\ref{eq:mlmt}
and~\ref{eq:scat} but with the parameters $\beta_0^{tm}$,
$\beta_1^{tm}$, $\beta_2^{tm}$, $\sigma_{tm}$ and $\sigma_{tm}'$
instead of those with index $\ell$. When $\beta_2^{\ell m}=0$ and
$\beta_2^{tm}=0$ we have ``self-similar'' evolution of the $L$--$M$
and $T$--$M$ relations respectively
\citep{Kaiser:86,Bryan:98}\footnote{Self-similar evolution is entirely
  determined by the $E(z)$ factors in the definitions of $\ell$, $t$
  and $m$.}. $\sigma_{\ell m}'=0$ and $\sigma_{tm}'=0$ correspond to
scaling relations with non-evolving scatter.

M10b showed that current data do not require departures from
self-similar evolution and constant scatter. R10 demonstrated that
$\gamma$ correlates weakly with departures from self-similarity and
constant scatter in the $L$--$M$ relation and negligibly for those in
the $T$--$M$ relation. Here we therefore assume self-similar evolution
and constant scatter for both relations ($\beta_2^{\ell
  m}=\sigma_{\ell m}'=\beta_2^{tm}=\sigma_{tm}'=0$).

\subsection{Galaxy clustering data}
\label{sec:gal}

For WiggleZ, a series of growth and expansion analyses have recently
been released, and here we build on one in particular: the joint
analysis of the AP effect and growth of structure presented by B11,
which contains four redshift bins of width $\Delta z = 0.2$, spanning
the redshift range $0.1 < z < 0.9$. The WiggleZ survey at the
Australian Astronomical Observatory was designed to extend the study
of large-scale structure over large cosmic volumes to higher redshifts
$z > 0.5$, complementing SDSS observations at lower redshifts. The
survey, which began in August 2006, completed observations in January
2011 and has obtained of order $200{,}000$ redshifts for UV-bright
emission-line galaxies covering of order 1000 square degrees of
equatorial sky.

For the WiggleZ analysis we fit our cosmological models to the joint
measurements of RSD and AP distortion presented by B11. For this, we
use the constraints obtained by B11 as a bivariate Gaussian likelihood
for $f \, \sigma_8(z)$ and $F(z)$, including the large correlations
between them. From B11, we have four bins with effective redshifts
$z=(0.22,0.41,0.60,0.78)$ and $f \,
\sigma_8(z)=(0.53\pm0.14,0.40\pm0.13,0.37\pm0.08,0.49\pm0.12)$,
$F(z)=(0.28\pm0.04,0.44\pm0.07,0.68\pm0.06,0.97\pm0.12)$ and
correlation coefficients $r=(0.83,0.94,0.89,0.84)$.

For the 6dFGS analysis we use the growth rate of structure measurement
obtained by \cite{Beutler:12}. The 6dFGS is a combined redshift and
peculiar velocity survey covering nearly the entire southern sky with
the exception of a $10$ degree band along the Galactic plane. Observed
galaxies were selected from the 2MASS Extended Source Catalog
\citep{Jarrett:00} and the redshifts were obtained with the 6-degree
Field multi-fibre instrument at the U.K.\ Schmidt Telescope between
2001 and 2006. The final 6dFGS sample contains about $125{,}000$
galaxies in $5$ bands distributed over $\sim 17{,}000$ square degrees
with a mean redshift of $z = 0.052$.

For the analysis of the RSD from 6dFGS data we use the constraints
obtained by \cite{Beutler:12} as a Gaussian likelihood for $f \,
\sigma_8(z) = 0.423 \pm 0.055$ at an effective redshift $z=0.067$.

The analysis of the SDSS-III BOSS results from \cite{Reid:12} are
based on the high-$z$ sample CMASS, which consists of $264{,}283$
galaxies in the redshift range $0.43<z<0.7$ over $3{,}275$ square
degrees. As part of SDSS-III \citep{Eisenstein:11}, BOSS has imaged
the South Galactic sky for an additional 3100 square degrees over
SDSS-II. This has increased the total sky coverage of SDSS imaging to
$14{,}055$ square degrees. As its primary goal, BOSS targets for
spectroscopy luminous galaxies selected from the SDSS imaging. Within
BOSS, CMASS is a roughly volume-limited sample of massive, luminous
galaxies \cite[for more detail see e.g.][]{Masters:11} tracing a
cosmological volume at a high enough density to enable powerful
statistical studies of large-scale structure.

For the analysis of the growth rate and AP effect measurements of
CMASS BOSS, we use a bivariate Gaussian likelihood for $f \,
\sigma_8(z)=0.43\pm0.07$ and $F(z)=0.68\pm0.04$ with a correlation
coefficient $r=0.87$ at an effective redshift $z=0.57$
\citep{Reid:12}\footnote{For the results in Section~\ref{sec:results}
  that include these and the distance-scale constraints from the BAO
  signature in the CMASS BOSS data, we extend this likelihood to
  account for the correlations between these three measurements as
  discussed in Section~\ref{sec:add-ons}.}. Note that this redshift is
similar to that of the third redshift bin of the WiggleZ analysis,
$z=0.6$. Due to the small overlap and the uncorrelated shot noise
between the two surveys, their covariance should be
minimal. Importantly, the results obtained by the two independent
experiments, which target very different galaxy types, and require
very specific studies of their nonlinear properties and modeling
uncertainties, are consistent.

\subsection{CMB data}
\label{sec:cmb}

For the CMB analysis, we use the data and likelihood
code\footnote{http://lambda.gsfc.nasa.gov/} from {\it
  WMAP}\footnote{Here we use the five-year {\it WMAP} data
  \citep[][and companion papers]{Dunkley:09, Komatsu:09}. For various
  of our results we have incorporated the galaxy data by importance
  sampling MCMC chains that had clusters and {\it WMAP}5 data. Note
  that, for $\Lambda$CDM and $w$CDM models, the constraints on
  $\Omega_{\rm m}h^2$, $\Omega_{\rm c}h^2$ (see comments on this
  parameter in Section~\ref{sec:results:gammaw}), $w$, $H_0$ and
  $\sigma_8$ from {\it WMAP}7 data alone do not differ significantly
  from those of {\it WMAP}5 (a maximum of 15 per cent in the errors of
  these parameters and much less in most cases), and even less when
  combined with additional data sets as those in
  Section~\ref{sec:add-ons}. We therefore expect a relatively small
  impact on the results from using {\it WMAP}7 instead. The new {\it
    WMAP}9 data, which appeared in the last stage of the present work,
  have up to 36 per cent better errors than {\it WMAP}5 for those
  parameters, and promises then a somewhat larger impact on the
  results.}. For the analyses including CMB data, we also fit for the
mean physical baryon and dark matter densities, $\Omega_{\rm b}h^2$
and $\Omega_{\rm c}h^2$, the optical depth to reionization, $\tau$,
the logarithm of the adiabatic scalar amplitude, $\ln(A_{\rm s})$,
which is related to $\sigma_8$, and the adiabatic scalar spectral
index, $n_{\rm s}$. For these analyses, instead of $H_0$ we fit
$\theta$, the (approximate) ratio of the sound horizon at last
scattering to the angular diameter distance, which is less correlated
with other parameters than $H_0$ \citep{Kosowsky:02}. We also
marginalize over the amplitude of the Sunyaev-Zel'dovich effect from
galaxy clusters, $0<A_{\rm SZ}< 2$ \citep{Spergel:07}.

\subsection{Additional data sets}
\label{sec:add-ons}

We also present results including constraints from the Union II SNIa
data set of \cite{Suzuki:12}, the SH0ES program of \cite{Riess:11},
and the BAO analyses of \cite{Percival:09}, at two intermediate
redshifts, and \cite{Reid:12}, at a higher redshift.

The SNIa data set consists of a compilation of 580 SNIa from a variety
of sources. For the likelihood analysis of these data we use the {\sc
  CosmoMC} module\footnote{http://supernova.lbl.gov/Union/} of
\cite{Suzuki:12}, including their treatment of the systematic
errors.

\begin{figure*}
\begin{center}
\includegraphics[width=3.25in]{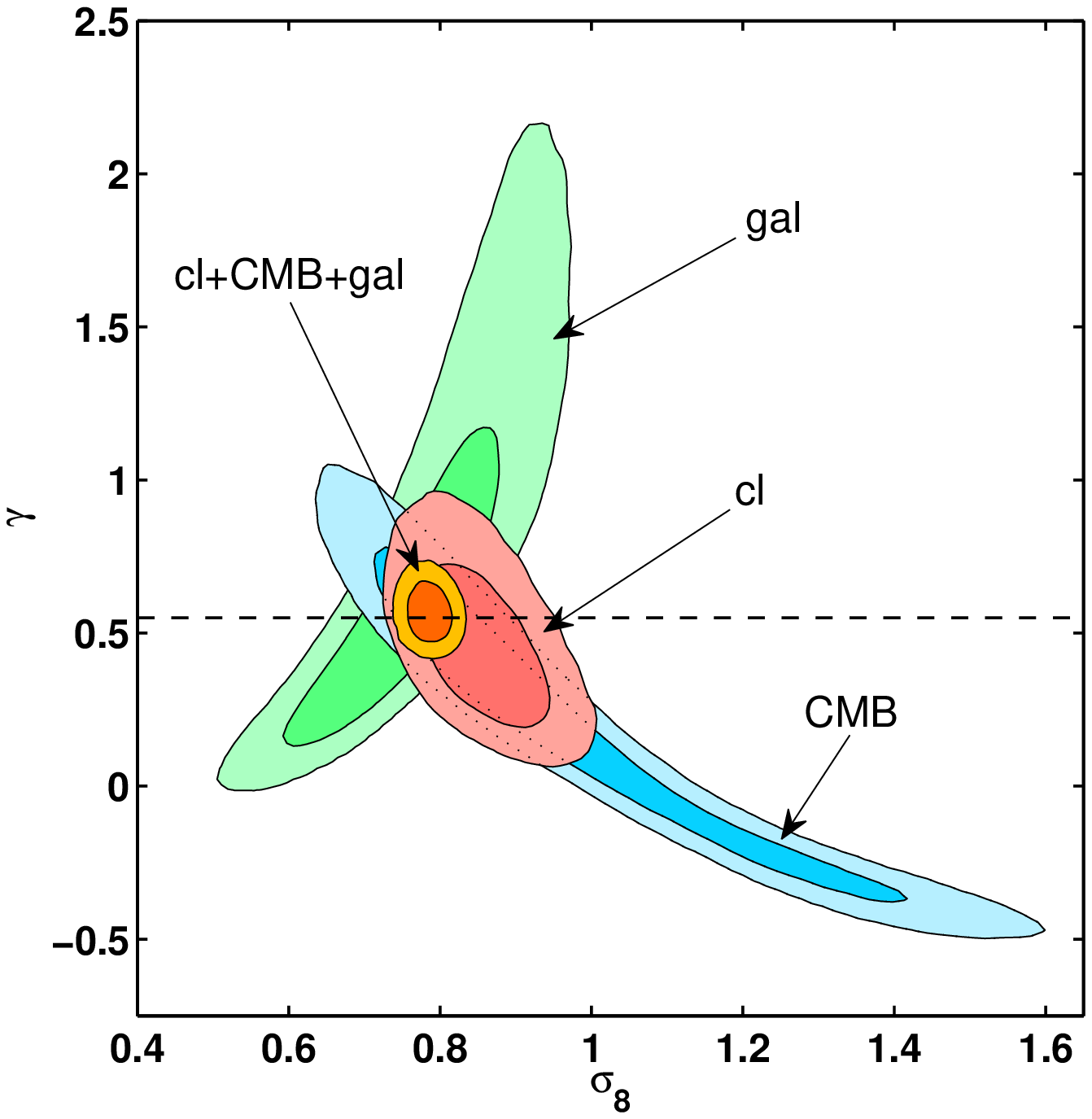}
\includegraphics[width=3.25in]{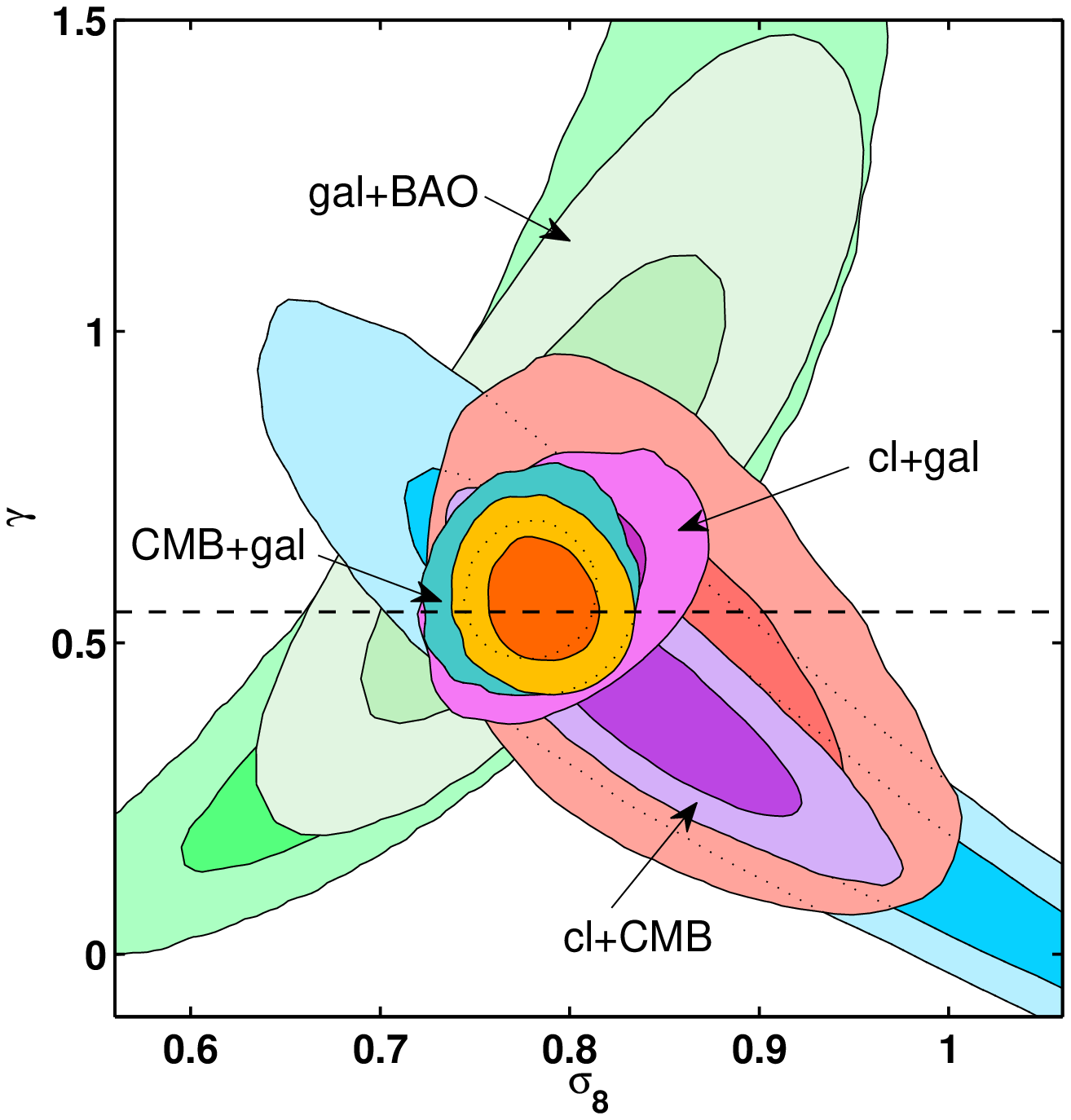}
\caption{68.3 and 95.4 per cent confidence contours in the $\sigma_8,
  \gamma$ plane for the $\gamma$+$\Lambda$CDM model. In the left
  panel, results are shown for the following individual data sets:
  galaxy growth (gal; green contours), CMB (blue contours) and cluster
  growth (cl; red contours). In the right panel, the same results are
  shown (for comparison purposes) together with the following
  combinations: cl+CMB (purple contours), cl+gal (magenta contours)
  and CMB+gal (turquoise contours). The combination of the three data
  sets breaks the degeneracies between $\gamma$ and $\sigma_8$ and
  provides tight constraints on this plane (inner, gold contours). In
  the right panel, results are also shown for the combination gal+BAO
  (pale green contours). The horizontal, dashed lines mark
  $\gamma=0.55$ (GR).}
\label{fig:datasets}
\end{center}
\end{figure*}

For the BAO analysis, we use the results and methods of
\cite{Percival:09}, based on 2dFGRS \citep{Colless:03} and SDSS-II DR7
\citep{Abazajian:09} data, by including bivariate Gaussian constraints
on the ratio $d_{\rm z}\equiv r_{\rm s}(z_{\rm d})/D_{\rm v}(z)$ at
$z=0.2$ and $z=0.35$, with the corresponding covariance between
$d_{0.2}$ and $d_{0.35}$, where $r_{\rm s}(z_{\rm d})$ is the sound
horizon at the drag epoch\footnote{To calculate $z_{\rm d}$ we use the
  exact expression \cite[see e.g. appendix B of][]{Hamann:10} instead
  of the approximate fitting formula of \cite{Eisenstein:98}. Note,
  however, that since \cite{Percival:09} and \cite{Reid:12} used the
  latter formula to fit the data, our results of the sound horizon
  need to be appropriately rescaled \cite[see
  again][]{Hamann:10}. Besides being more accurate, the exact
  calculation of $z_{\rm d}$ is independent of the standard
  assumptions used to obtain the fitting formula, and therefore valid
  for other models. Interestingly, though, for the extended models
  used here we find no significant differences in the results obtained
  from using either calculation.}  and $D_{\rm
  v}(z)\equiv[(1+z)^{2}D_{\rm
  A}(z)^{2}c\,z/H(z)]^{1/3}$. \cite{Percival:09} showed that these
results can also be recast as approximately independent Gaussian
constraints on $d_{0.275}$ and the ratio of the distance scales
$D_{\rm v}(0.35)/D_{\rm v}(0.2)$. For the BAO results of
\cite{Reid:12}, based on CMASS BOSS data, we extend the bivariate
Gaussian likelihood of Section~\ref{sec:gal} to a trivariate Gaussian
likelihood by including a constraint on $\alpha\equiv[d_{\rm z}]_{\rm
  fiducial}/d_{\rm z}=1.023\pm0.019$, at $z=0.57$, and the
corresponding correlation coefficients $r_{f \sigma_8 \alpha}=-0.0086$
and $r_{\rm F \alpha}=-0.080$ \cite[see Section 6.4 of][]{Reid:12}.

Note that the overlap between the ranges in redshift of the SDSS-II
DR7 luminous red galaxy sample ($0.16<z<0.47$) used in the
\cite{Percival:09} results and the SDSS-III DR9 CMASS sample
($0.43<z<0.7$) used in the BOSS results \citep{Reid:12} is very small
\citep[see this comparison e.g. in][]{Anderson:12}. Therefore, we
assume that the two BAO measurements are essentially independent and
can be straightforwardly combined.

For the SH0ES analysis, we use a Gaussian prior on
$H_0=73.8\pm2.4\Hunit$. This measurement is based on {\it Hubble Space
  Telescope} optical and infrared data for over $600$ Cepheid
variables in the host galaxies of $8$ nearby SNIa \citep{Riess:11}.

\begin{table*}
\begin{center}
  \caption{Marginalized mean values and 68.3 per cent confidence
    limits for the $\gamma$+$\Lambda$CDM and $\gamma$+$w$CDM models
    using various subsets of the data. When the CMB data is not
    included the BBNS and SH0ES priors are used. For gal+BAO and
    $\gamma$+$\Lambda$CDM, a slightly tighter constraint than the
    prior on $H_0$ is obtained.}
\label{table:params}
\begin{tabular}{ c c c c c c}
\hline
   Data                             & $\Omega_{\rm m}$                      & $H_0(\Hunit)$                       & $w$                              & $\sigma_8$                         &  $\gamma$ \\
\hline

\noalign{\vskip 4pt}

cl                                  & $ 0.216^{+0.032}_{-0.032} $     &     $ 73.7^{+2.4}_{-2.4} $          & -1                                 &     $ 0.856^{+0.054}_{-0.055} $     &     $ 0.469^{+0.175}_{-0.177} $     \\        
\noalign{\vskip 4pt}

cl+CMB                              & $ 0.249^{+0.021}_{-0.021} $     &     $ 72.5^{+2.1}_{-2.1} $          & -1                                 &     $ 0.844^{+0.049}_{-0.049} $     &     $ 0.415^{+0.128}_{-0.126} $     \\    
\noalign{\vskip 4pt}

cl+gal                              & $ 0.236^{+0.033}_{-0.034} $     &     $ 73.1^{+2.4}_{-2.4} $          & -1                                 &     $ 0.795^{+0.030}_{-0.030} $     &     $ 0.586^{+0.090}_{-0.089} $     \\    
\noalign{\vskip 4pt}

gal+BAO                             & $ 0.366^{+0.049}_{-0.049} $     &     $ 72.8^{+2.3}_{-2.3} $          & -1                                 &     $ 0.795^{+0.063}_{-0.063} $     &     $ 0.780^{+0.255}_{-0.256} $     \\ 
\noalign{\vskip 4pt}

CMB+gal                             & $ 0.256^{+0.027}_{-0.027} $     &     $ 71.9^{+2.4}_{-2.4} $          & -1                                 &     $ 0.779^{+0.023}_{-0.023} $     &     $ 0.588^{+0.073}_{-0.073} $     \\    
\noalign{\vskip 4pt}

cl+CMB+gal                          & $ 0.254^{+0.022}_{-0.022} $     &     $ 72.2^{+2.0}_{-2.1} $          & -1                                 &     $ 0.785^{+0.019}_{-0.019} $     &     $ 0.570^{+0.064}_{-0.063} $     \\    
\noalign{\vskip 4pt}

cl+CMB+gal+BAO                      & $ 0.284^{+0.013}_{-0.013} $     &     $ 69.4^{+1.1}_{-1.1} $          & -1                                 &     $ 0.789^{+0.019}_{-0.019} $     &     $ 0.618^{+0.063}_{-0.063} $     \\ 
\noalign{\vskip 4pt}

cl+CMB+gal+SH0ES                    & $ 0.247^{+0.017}_{-0.017} $     &     $ 72.8^{+1.5}_{-1.6} $          & -1                                 &     $ 0.784^{+0.019}_{-0.019} $     &     $ 0.561^{+0.061}_{-0.061} $     \\ 
\noalign{\vskip 4pt}

cl+CMB+gal+SNIa                     & $ 0.264^{+0.019}_{-0.020} $     &     $ 71.3^{+1.8}_{-1.8} $          & -1                                 &     $ 0.788^{+0.019}_{-0.019} $     &     $ 0.585^{+0.065}_{-0.064} $     \\ 
\noalign{\vskip 4pt}

cl+CMB+gal+SNIa+BAO                 & $ 0.284^{+0.012}_{-0.012} $     &     $ 69.4^{+1.1}_{-1.1} $          & -1                                 &     $ 0.790^{+0.019}_{-0.019} $     &     $ 0.618^{+0.062}_{-0.062} $     \\ 
\noalign{\vskip 4pt}

cl+CMB+gal+SNIa+SH0ES               & $ 0.255^{+0.015}_{-0.016} $     &     $ 72.1^{+1.5}_{-1.4} $          & -1                                 &     $ 0.786^{+0.019}_{-0.019} $     &     $ 0.575^{+0.060}_{-0.060} $     \\ 
\noalign{\vskip 4pt}

cl+CMB+gal+SNIa+SH0ES+BAO           & $ 0.277^{+0.011}_{-0.011} $     &     $ 70.2^{+1.0}_{-1.0} $         & -1                                  &    $ 0.791^{+0.019}_{-0.019} $     &     $ 0.616^{+0.061}_{-0.061} $     \\ 
\noalign{\vskip 4pt}

cl                                  & $ 0.220^{+0.034}_{-0.034} $     &     $ 73.7^{+2.4}_{-2.4} $          &     $ -1.021^{+0.190}_{-0.187} $      &     $ 0.855^{+0.056}_{-0.057} $     &     $ 0.507^{+0.236}_{-0.242} $     \\                                                                                                                                                                                                               
\noalign{\vskip 4pt}
                             
cl+CMB                              & $ 0.240^{+0.030}_{-0.030} $     &     $ 74.4^{+4.4}_{-4.4} $          &     $ -1.062^{+0.122}_{-0.122} $     &     $ 0.851^{+0.052}_{-0.052} $     &     $ 0.454^{+0.148}_{-0.149} $     \\  
\noalign{\vskip 4pt}

cl+gal                              & $ 0.216^{+0.036}_{-0.035} $     &     $ 73.5^{+2.4}_{-2.4} $          &     $ -0.865^{+0.109}_{-0.107} $     &     $ 0.812^{+0.035}_{-0.035} $     &     $ 0.502^{+0.102}_{-0.103} $     \\  
\noalign{\vskip 4pt}

gal+BAO                             & $ 0.323^{+0.060}_{-0.060} $     &     $ 73.9^{+2.4}_{-2.4} $     &     $ -1.199^{+0.176}_{-0.179} $     &     $ 0.721^{+0.081}_{-0.085} $     &     $ 0.616^{+0.265}_{-0.266} $     \\ 
\noalign{\vskip 4pt}

CMB+gal                             & $ 0.278^{+0.034}_{-0.034} $     &     $ 69.0^{+3.4}_{-3.4} $          &     $ -0.909^{+0.081}_{-0.082} $     &     $ 0.772^{+0.023}_{-0.023} $     &     $ 0.526^{+0.088}_{-0.088} $     \\  
\noalign{\vskip 4pt}

cl+CMB+gal                          & $ 0.263^{+0.024}_{-0.025} $     &     $ 70.6^{+2.7}_{-2.7} $          &     $ -0.950^{+0.069}_{-0.070} $     &     $ 0.780^{+0.020}_{-0.020} $     &     $ 0.533^{+0.080}_{-0.080} $     \\  
\noalign{\vskip 4pt}

cl+CMB+gal+BAO                      & $ 0.289^{+0.015}_{-0.015} $     &     $ 68.3^{+1.8}_{-1.8} $     &     $ -0.941^{+0.073}_{-0.073} $     &     $ 0.783^{+0.020}_{-0.020} $     &     $ 0.562^{+0.086}_{-0.087} $     \\ 
\noalign{\vskip 4pt}

cl+CMB+gal+SH0ES                    & $ 0.248^{+0.017}_{-0.017} $     &     $ 72.4^{+1.8}_{-1.9} $          &     $ -0.983^{+0.060}_{-0.059} $     &     $ 0.783^{+0.020}_{-0.020} $     &     $ 0.549^{+0.080}_{-0.080} $     \\  
\noalign{\vskip 4pt}

cl+CMB+gal+SNIa                     & $ 0.267^{+0.019}_{-0.019} $     &     $ 70.1^{+1.9}_{-1.9} $          &     $ -0.939^{+0.053}_{-0.053} $     &     $ 0.780^{+0.020}_{-0.020} $     &     $ 0.530^{+0.077}_{-0.076} $     \\  
\noalign{\vskip 4pt}

cl+CMB+gal+SNIa+BAO                 & $ 0.288^{+0.013}_{-0.013} $     &     $ 68.6^{+1.4}_{-1.4} $     &     $ -0.950^{+0.055}_{-0.056} $     &     $ 0.784^{+0.019}_{-0.019} $     &     $ 0.569^{+0.079}_{-0.079} $     \\ 
\noalign{\vskip 4pt}

cl+CMB+gal+SNIa+SH0ES               & $ 0.255^{+0.015}_{-0.015} $     &     $ 71.6^{+1.5}_{-1.5} $     &     $ -0.961^{+0.050}_{-0.051} $     &     $ 0.782^{+0.020}_{-0.020} $     &     $ 0.537^{+0.077}_{-0.076} $     \\ 
\noalign{\vskip 4pt}

cl+CMB+gal+SNIa+SH0ES+BAO           & $ 0.278^{+0.012}_{-0.011} $     &     $ 70.0^{+1.3}_{-1.3} $     &     $ -0.987^{+0.054}_{-0.053} $     &     $ 0.789^{+0.019}_{-0.019} $     &     $ 0.604^{+0.078}_{-0.078} $     \\ 

\hline

\end{tabular}
\end{center}
\end{table*}

\subsection{MCMC implementation}
\label{sec:data}

To calculate the parameter posterior probability distribution
functions (pdf's) we use the Metropolis-Hastings Markov Chain Monte
Carlo (MCMC) algorithm, as implemented in the code {\sc
  CosmoMC}\footnote{http://cosmologist.info/cosmomc/}~\citep{Lewis:02}. We
employ a modified version of this code that includes additional
modules for the likelihood analyses of the cluster growth experiment
(M10a) and the $f_{\rm gas}$ experiment \citep{Rapetti:05,
  Allen:08}\footnote{http://www.slac.stanford.edu/$\sim$drapetti/fgas\_module/}. In
this version of the code, we have also incorporated the RSD, AP effect
and BAO (BOSS) analyses as {\sc CosmoMC} modules\footnote{We use these
  and standard modules for the other data sets in
  Section~\ref{sec:add-ons} either to run new MCMC chains or to
  perform importance sampling on existing ones. For selected examples,
  we have explicitly checked that both methods provide the same
  results.}. We also use a modified version of the code {\sc
  camb}\footnote{http://camb.info/}~\citep{Lewis:00} that includes
$\gamma$ in the analysis of the ISW effect of the CMB data
\citep{Rapetti:09}.

For our most general model, we simultaneously fit a total of $34$
parameters. From these, $8$ are cosmological parameters and $26$ are
used to model astrophysical variables and marginalize over systematic
uncertainties: $1$ for CMB (see Section~\ref{sec:cmb}), $7$ for
$f_{\rm gas}$ \cite[see details in][]{Allen:08} and $18$ for cluster
growth data (see Sections~\ref{sec:cosmo:growth},~\ref{sec:scal}, and
M10a,b for full details).

For analyses without CMB data, we fix $n_{\rm s}$ to $0.95$ since, for
such analyses, $n_{\rm s}$ is degenerate with $\sigma_{8}$ (see M10a).
For these analyses, we also use Gaussian priors on $H_0$ from the
SH0ES program \citep{Riess:11}, and $\Omega_{\rm b}
h^2=0.0213\pm0.0010$ from Big Bang Nucleosynthesis (BBNS) studies
\citep{Pettini:08}.

\begin{figure*}
\begin{center}
\includegraphics[width=3.25in]{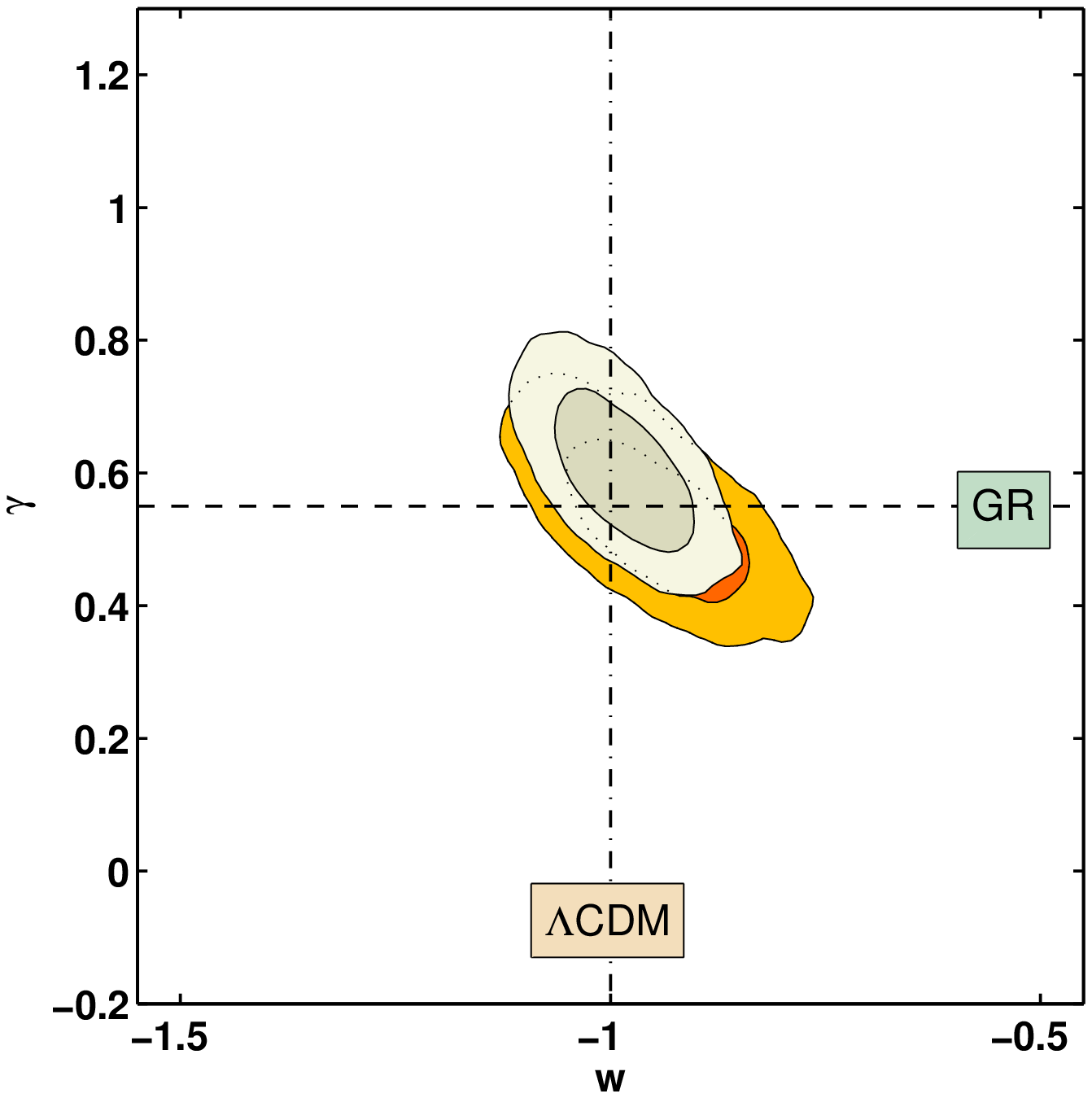}
\includegraphics[width=3.25in]{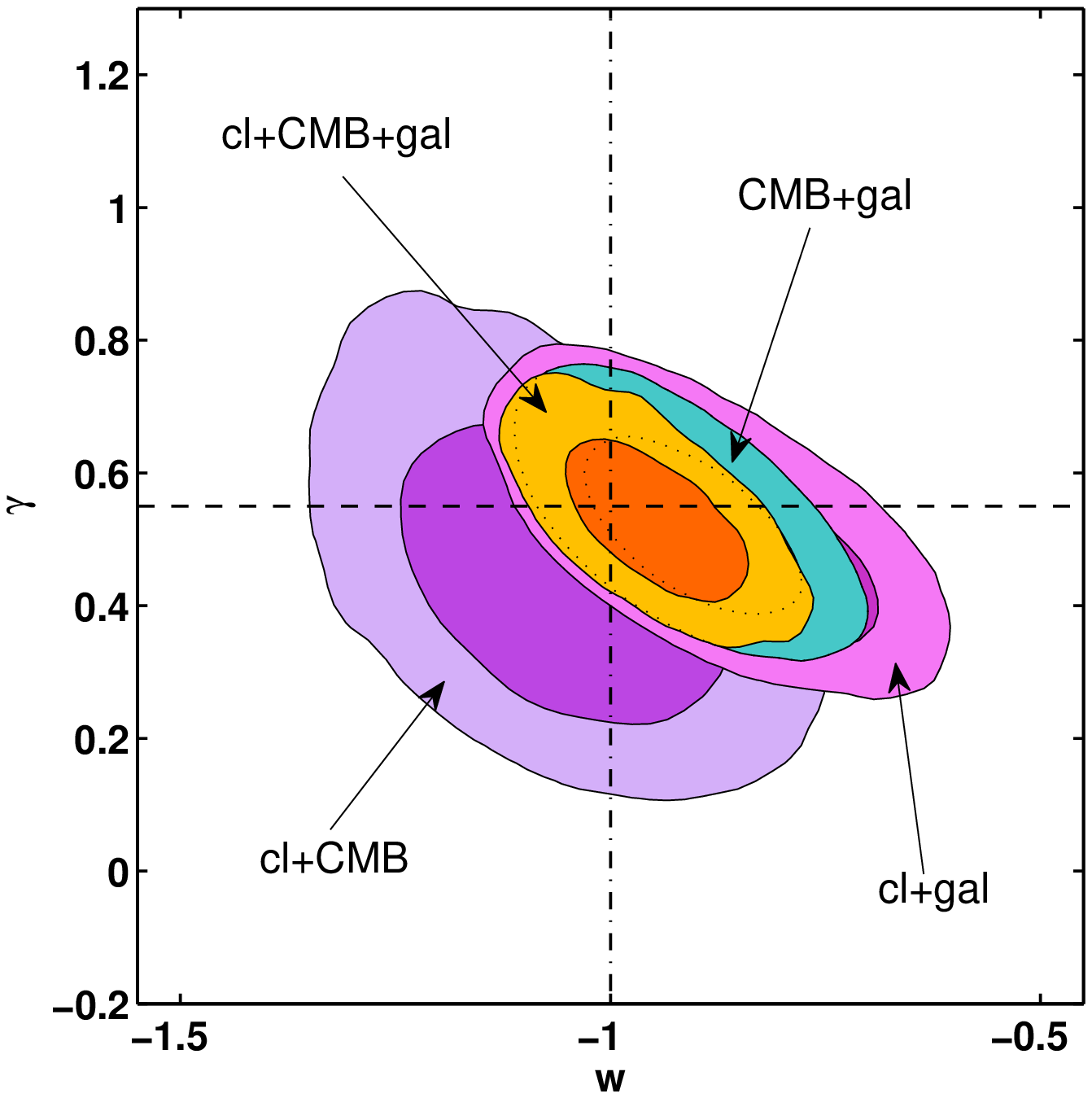}
\caption{68.3 and 95.4 per cent confidence contours in the $w,\gamma$
  plane for the $\gamma$+$w$CDM model from cl+CMB+gal (gold contours
  in both panels) and the following combinations of data (right
  panel): cl+CMB (purple contours), cl+gal (magenta contours), and
  CMB+gal (turquoise contours). The platinum contours in the left
  panel correspond to adding SNIa+SH0ES+BAO to the combination of
  cl+CMB+gal. The horizontal, dashed lines mark $\gamma=0.55$, the
  growth history for GR. The vertical, dot-dashed lines mark $w=-1$,
  the expansion history for $\Lambda$CDM. This figure shows that the
  results are simultaneously consistent with GR and $\Lambda$CDM.}
\label{fig:wconst}
\end{center}
\end{figure*}

\section{Results}
\label{sec:results}

\subsection{Constraints on the $\gamma$+$\Lambda$CDM model}
\label{sec:results:gammalcdm}

The left panel of Figure~\ref{fig:datasets} shows the joint
constraints in the $\sigma_8, \gamma$ plane for the
$\gamma$+$\Lambda$CDM model. The green contours show the constraints
obtained from the RSD and AP effect data from WiggleZ, 6dFGS and CMASS
BOSS (hereafter referred as galaxy/gal data); the blue contours those
from the CMB data; and the red contours those from the cluster
abundance and $f_{\rm gas}$ data (hereafter referred as cluster/cl
data). Combining the cluster+CMB+galaxy data we obtain the tight
constraints shown by the gold contours.

\begin{figure*}
\begin{center}
\includegraphics[width=3.25in]{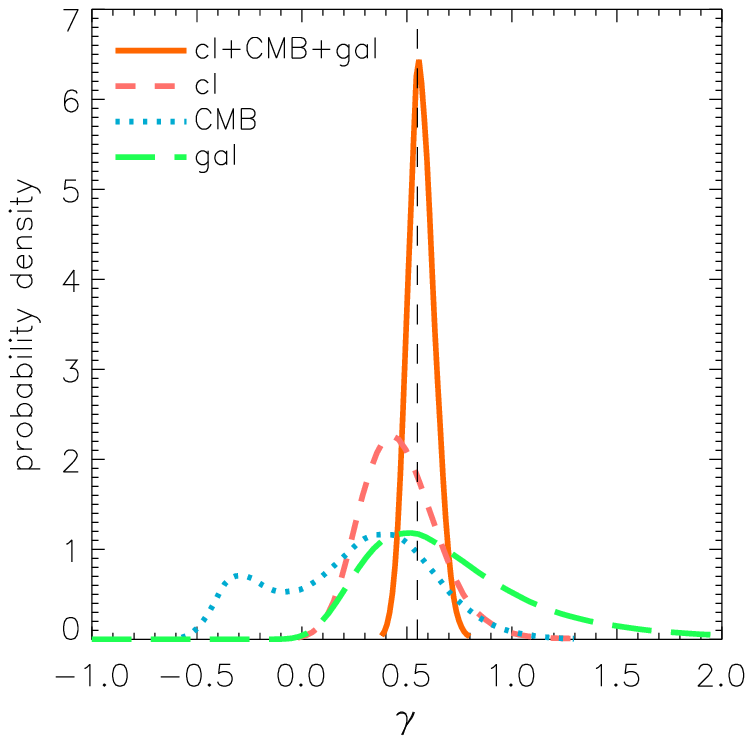}
\includegraphics[width=3.25in]{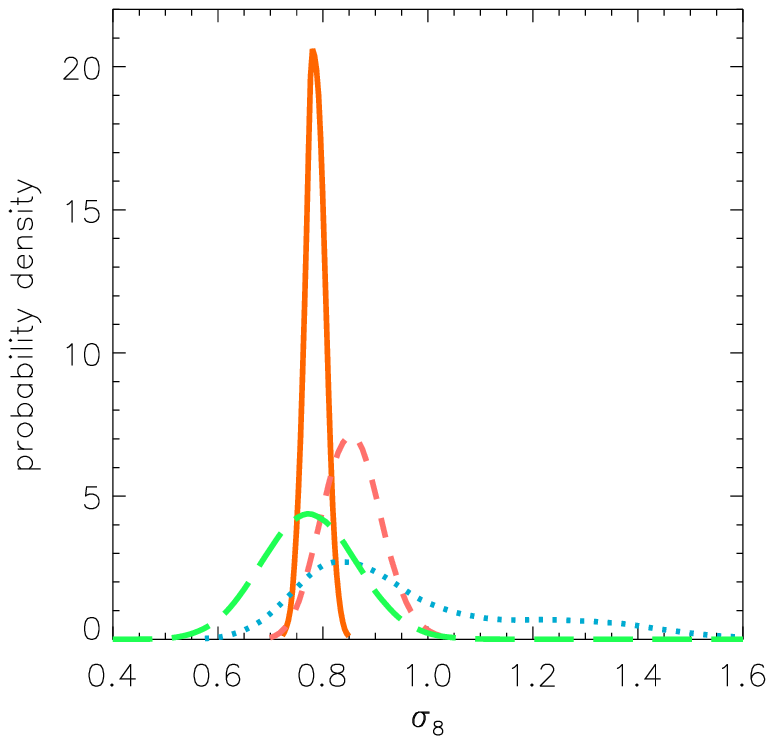}
\caption{Marginalized probability distribution functions for $\gamma$
  (left panel) and $\sigma_8$ (right panel) for the
  $\gamma$+$\Lambda$CDM model. Results are shown for the following
  data sets: CMB (blue, dotted line), gal (green, long-dashed line),
  cl (red, dashed line) and cl+CMB+gal (gold, solid line). In the left
  panel, the vertical, dashed line marks $\gamma=0.55$ (GR).}
\label{fig:1dlcdm}
\end{center}
\end{figure*}

As shown in the figure, individually, the CMB and galaxy data exhibit
significant degeneracies in the $\sigma_8, \gamma$ plane, as expected
(see Section~\ref{sec:physics}). For the cluster data, the correlation
between these two parameters is much weaker, enabling independent
constraints on both parameters.\footnote{When combining gal+BAO we
  also obtain results that are more comparable to those of the
  clusters due to the degeneracies broken by this combination (see
  further details in the text).} Importantly, the constraints from the
three independent experiments (which are affected by very different
systematic uncertainties) are in excellent agreement. This agreement
motivates us to combine the constraints, leading to the results shown
in the inner, gold contours. Combining the three data sets we obtain
marginalized constraints on $\gamma=0.570^{+0.064}_{-0.063}$ (in good
agreement with GR) and $\sigma_8=0.785\pm0.019$ (see also
Table~\ref{table:params}).\footnote{For this paper, we quote
  marginalized mean values and central credible intervals. For the
  latter, equal fractions of the volume of the posterior lie on each
  side of the interval \cite[see e.g.][]{Hamann:07}. Instead, in R10
  results were presented using marginalized peak values and minimal
  credible intervals, for which the size of the interval is
  minimized. For approximately symmetric posteriors, such as those for
  our combined data, both choices provide similar results, although by
  construction those from the former tend to be slightly more
  conservative. For our individual data sets, for which the posteriors
  are less symmetric, we show full marginalized distributions in
  Figures~\ref{fig:1dlcdm} and~\ref{fig:1dwcdm}.}

If we also include SNIa, BAO and the SH0ES measurement of $H_0$, the
constraints on the growth parameters are, as expected, almost the same
(see Table~\ref{table:params}) although interestingly we obtain a
small $4$ per cent improvement in the error in $\gamma$. For this
combination we also obtain improved, tight constraints on the
expansion parameters $\Omega_{\rm m}=0.277\pm0.011$ and
$H_0=70.2\pm1.0\Hunit$. It is worth noting that the addition of the
BAO data alone provides almost the same improved constraints on the
$\Omega_{\rm m}, H_0$ plane as those from adding all three data sets
(see Table~\ref{table:params}).

The right panel of Figure~\ref{fig:datasets} shows a zoom into the
central regions of the constraints shown in the left panel, together
with the constraints for the combinations of cl+CMB data (purple
contours), cl+gal data (magenta contours) and CMB+gal data (turquoise
contours). The gold, tightest contours correspond again to the
combination of the three data sets, cl+CMB+gal. Notably, the nearly
orthogonal degeneracies of the CMB (blue contours) and galaxy (green
contours) constraints allow their combination (turquoise contours) to
provide tight marginalized constraints in the growth plane. The area
enclosed by the 95.4 per cent confidence contour in the $\sigma_8,
\gamma$ plane is only slightly more than one third larger for CMB+gal
than for the three data sets combined.

As found by R10\footnote{Note that the results presented in R10 were
  for a combination of cluster+CMB+SNIa+BAO data. However, the
  constraints on the $\sigma_8, \gamma$ plane were primarily driven by
  the cluster+CMB data.}, for the cl+CMB data (purple contours)
$\sigma_8$ and $\gamma$ are highly correlated, with a correlation
coefficient $\rho=-0.85$. The addition of the galaxy data breaks this
degeneracy. With respect to constraints obtained from cl+CMB, those
for the cl+CMB+gal provide more than a factor $2$ reduction in the
area enclosed by the 95.4 per cent confidence contour in the
$\sigma_8, \gamma$ plane.

In the right panel of the figure, we also show the constraints from
the combination gal+BAO (pale green contours), for which both data
sets come from the analysis of different properties of galaxy redshift
surveys. Interestingly, even though the baryon acoustic oscillation
data on their own provide only constraints on expansion parameters,
those on $\Omega_{\rm m}$ help in reducing the large degeneracies that
the galaxy growth data has in the $\Omega_{\rm m}, \gamma$ and
$\Omega_{\rm m}, \sigma_8$ planes, with correlation coefficients of
$\rho=0.83$ and $\rho=0.74$ for each plane. Adding BAO to gal we
obtain then a significant improvement in the constraints on the growth
plane $\sigma_8, \gamma$.

\begin{figure*}
\begin{center}
\includegraphics[width=2.3in]{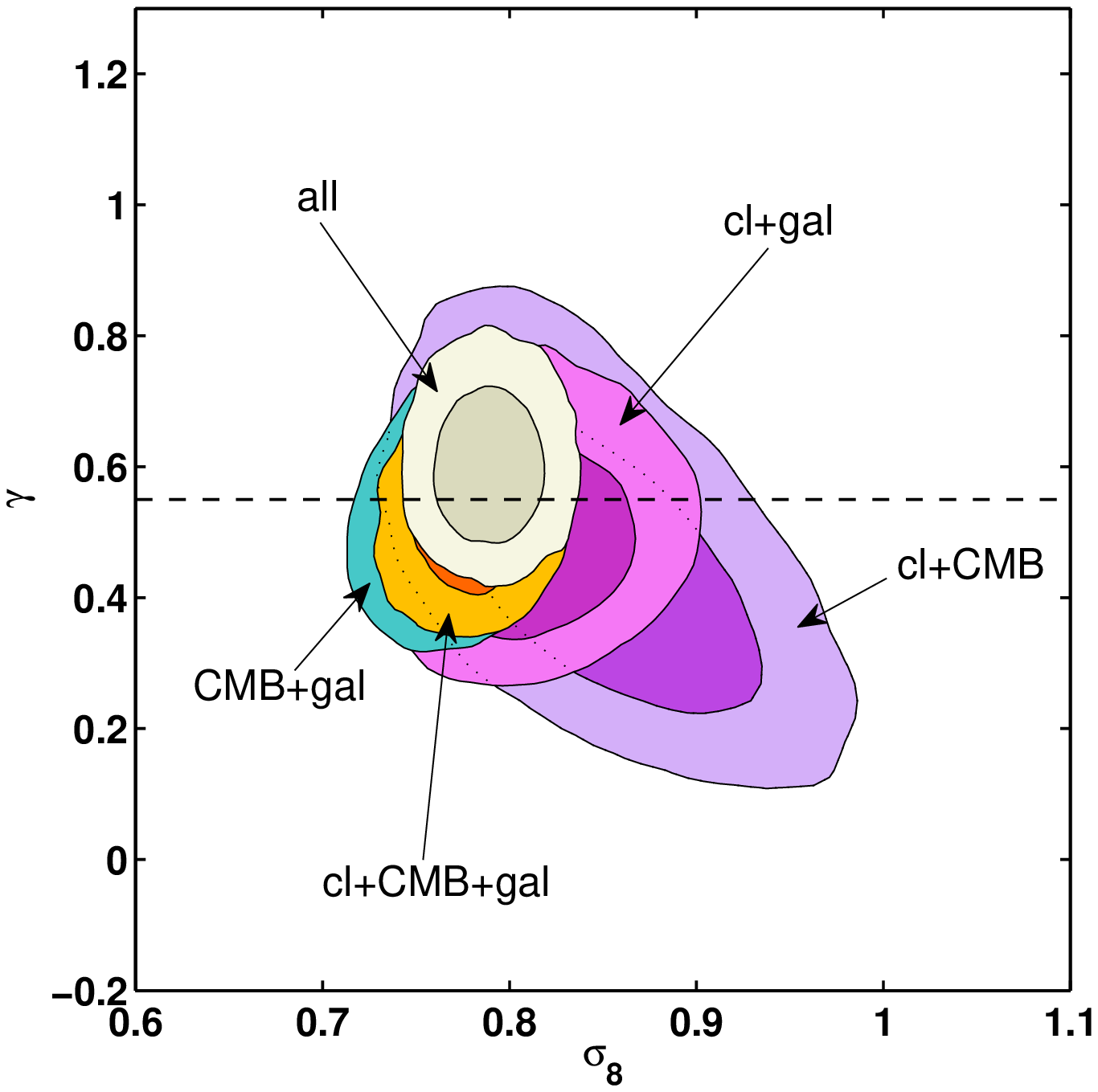}
\includegraphics[width=2.3in]{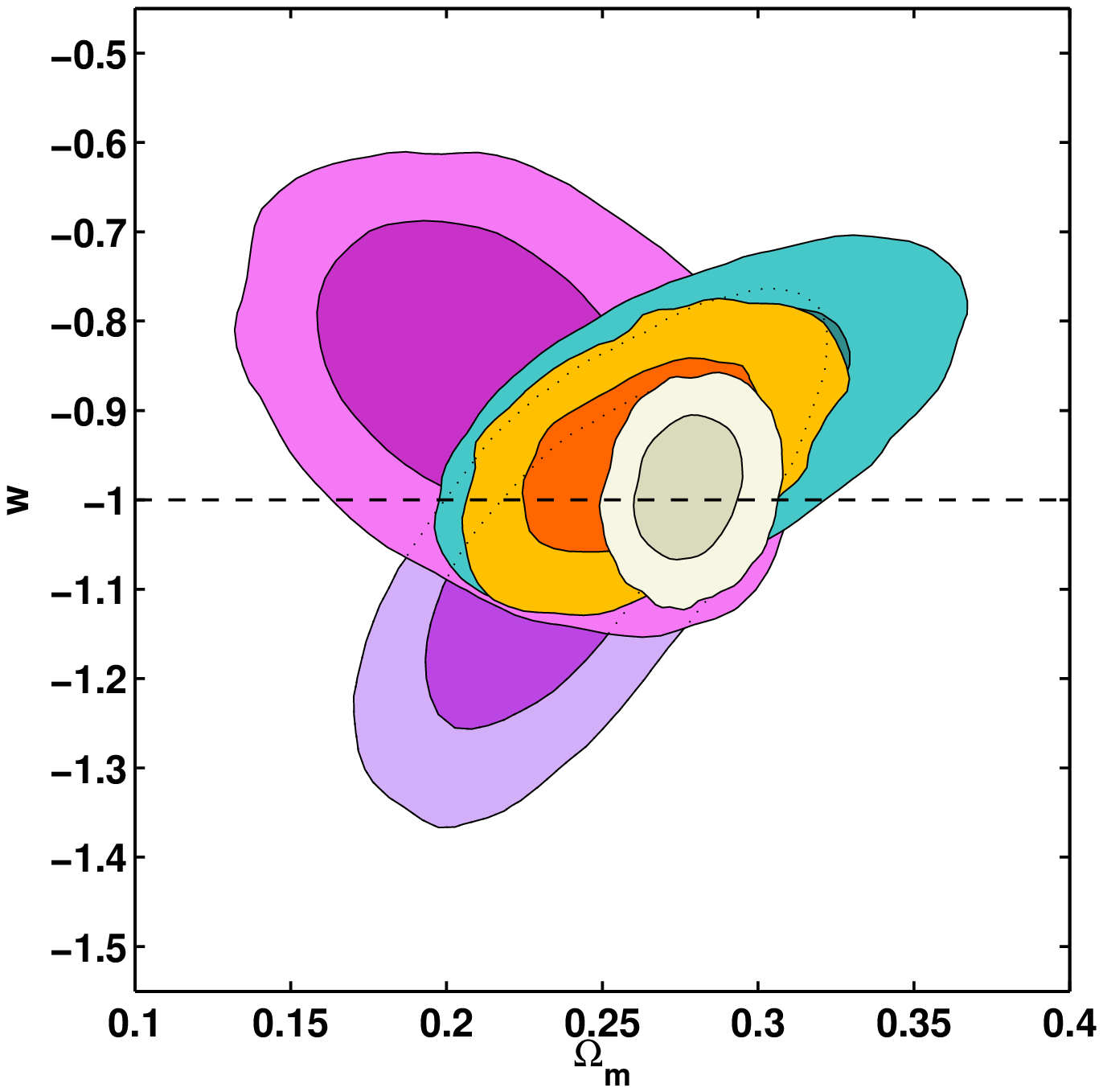}
\includegraphics[width=2.27in]{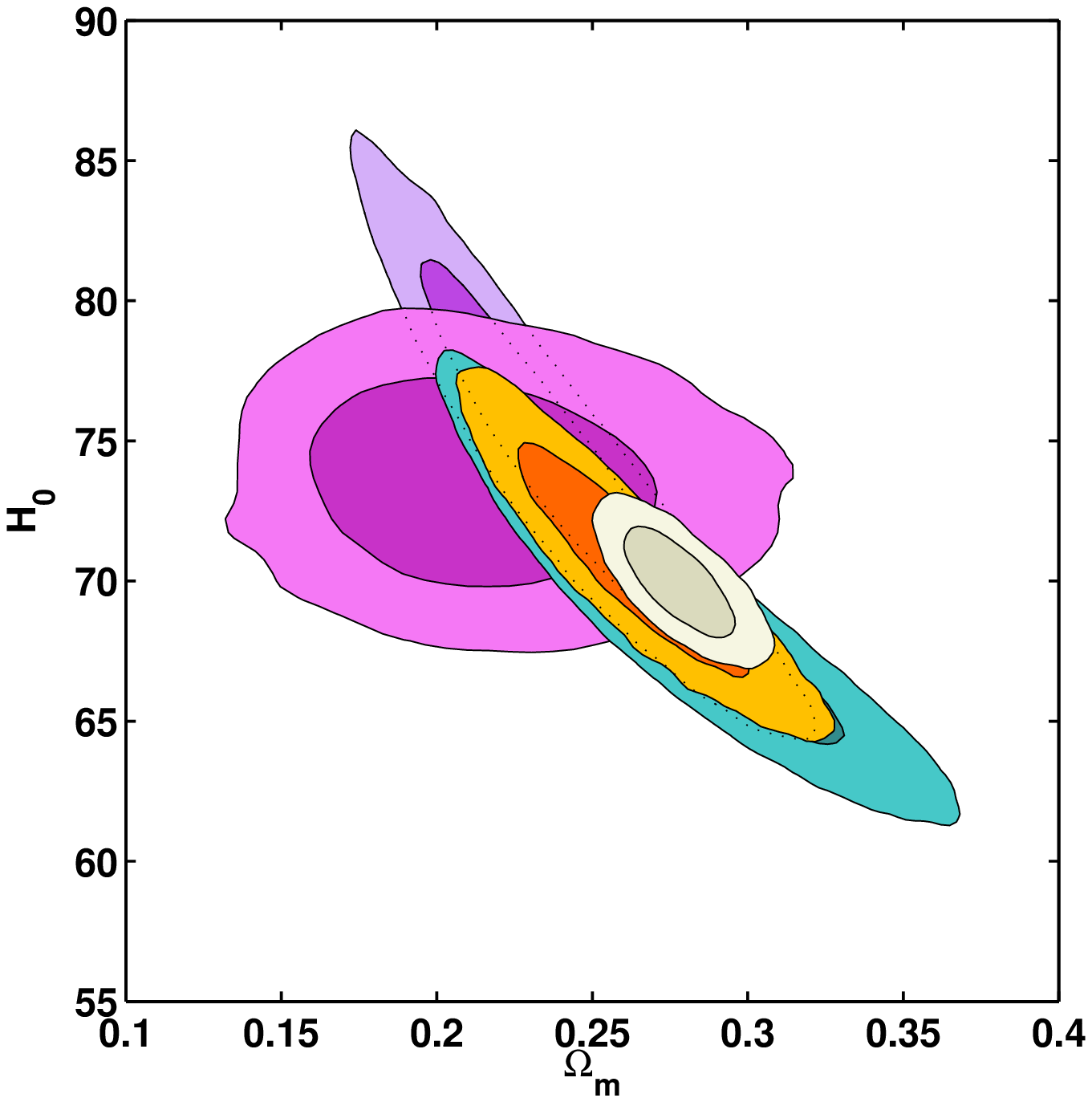}
\caption{68.3 and 95.4 per cent confidence contours in the growth
  plane $\sigma_8,\gamma$ (left panel) and the expansion planes
  $\Omega_{\rm m}, w$ (middle panel) and $\Omega_{\rm m}, H_0$ (right
  panel) for the $\gamma$+$w$CDM model. The three panels show
  constraints from the following combinations of data sets: cl+CMB
  (purple contours), cl+gal (magenta contours), CMB+gal (turquoise
  contours), and cl+CMB+gal (gold contours). The platinum contours
  correspond to adding SNIa+SH0ES+BAO to the latter combination of
  data. The horizontal, dashed lines mark $\gamma=0.55$ (GR; left
  panel) and $w=-1$ ($\Lambda$CDM; middle panel).}
\label{fig:wconstplanes}
\end{center}
\end{figure*}

\subsection{Constraints on the $\gamma$+$w$CDM model}
\label{sec:results:gammaw}

The left panel of Figure~\ref{fig:wconst} shows the joint constraints
in the $w, \gamma$ plane for the $\gamma$+$w$CDM model. For the
combination of our primary data sets, cl+CMB+gal, we obtain the gold
contours. For these, we find marginalized constraints on
$w=-0.950^{+0.069}_{-0.070}$ and $\gamma=0.533\pm0.080$ at the 68.3
per cent confidence level. These results are simultaneously consistent
with GR and $\Lambda$CDM. The platinum contours in this panel show the
joint constraints on the $w, \gamma$ plane when we add SNIa+SH0ES+BAO
to the cl+CMB+gal data\footnote{From Table~\ref{table:params}, note
  that the main additional constraint on this plane comes from the
  SNIa data. The SH0ES and BAO data, though, significantly help in
  constraining the combination of parameters $\Omega_{\rm m}$ and
  $H_0$.}. In this case, we find marginalized constraints of
$w=-0.987^{+0.054}_{-0.053}$ and $\gamma=0.604\pm0.078$. Again the
results are consistent with GR+$\Lambda$CDM.

In the right panel of the figure, the purple contours correspond to
cl+CMB, the magenta contours to cl+gal, the turquoise contours to
CMB+gal and the gold contours again to the combination of the three
data sets. The horizontal, dashed and vertical, dot-dashed lines mark
$\gamma=0.55$ (GR) and $w=-1$ ($\Lambda$CDM), respectively.

Comparing the cl+CMB with the cl+CMB+gal results, we find $46$ and
$62$ per cent improvements in the constraints on $\gamma$ and
$\sigma_8$. It is also worth noting that the improvement in the joint
measurement of $w$ and $\gamma$ is larger than that for each
individual parameter. We find more than a factor $3$ reduction in the
area enclosed by the 95.4 per cent confidence contour of the joint $w,
\gamma$ constraints. Note that the correlation between $w$ and
$\gamma$ increases from $\rho=-0.47$, for cl+CMB, to $\rho=-0.66$, for
cl+CMB+gal, which suggests that additional constraints on $w$ might
also help improving those on $\gamma$. In fact, even though SNIa and
SH0ES data provide direct additional constraints on only cosmic
expansion parameters, for which we obtain e.g. a $27$ per cent
improvement on $w$ when adding them to cl+CMB+gal, the combined,
marginalized constraints on $\gamma$ represent a small improvement of
$4$ per cent due to the correlation between $w$ and $\gamma$. For
these data sets combined,
cl+\allowbreak{}CMB+\allowbreak{}gal+\allowbreak{}SNIa+\allowbreak{}SH0ES,
the correlation in the $w, \gamma$ plane is still of
$\rho=-0.65$. Interestingly, the correlation between $\gamma$ and
$\Omega_{\rm c}h^2$ is also relatively large,
$\rho=0.72$,\footnote{The correlation between $\gamma$ and the CMB
  shift parameter, which is highly correlated with $\Omega_{\rm c}h^2$
  ($\rho=0.92$) and is purely an expansion parameter at high-$z$
  \cite[see the definition and further details in e.g.][]{Komatsu:09},
  is also large, $\rho=0.68$. The correlations between $\gamma$ and
  $\Omega_{\rm m}(=\Omega_{\rm b}+\Omega_{\rm c})$, $\rho=0.32$,
  $\gamma$ and $\Omega_{\rm b}h^2$, $\rho=0.11$, and $\gamma$ and
  $H_0$, $\rho=0.13$, are significantly smaller. As shown in the right
  panel of Figure~\ref{fig:wconstplanes} for other data combinations,
  the correlation between $\Omega_{\rm m}$ and $H_0$ is also large,
  $\rho=-0.82$.}  which indicates that e.g. the significant
improvements in the constraints on this parameter from the CMB
measurements of the {\it Planck}
satellite\footnote{http://www.esa.int/Planck} should help with
constraining $\gamma$.

Figure~\ref{fig:wconstplanes} shows constraints for the same model and
subsets of the data for three different planes: the growth plane
$\sigma_8, \gamma$ (left panel) and the expansion planes $\Omega_{\rm
  m}, w$ (middle panel) and $\Omega_{\rm m}, H_0$ (right panel). The
left panel of this figure shows that the correlation between
$\sigma_8$ and $\gamma$ reduces dramatically from cl+CMB (purple
contours), $\rho=-0.56$, to cl+CMB+gal (gold contours), a negligible
$\rho=0.08$. The reduction in the area enclosed by the 95.4 confidence
contour in this growth plane when adding gal to the cl+CMB data is
substantial.

The platinum contours in the three panels correspond to the
constraints obtained when adding
SNIa+\allowbreak{}SH0ES+\allowbreak{}BAO to the
cl+\allowbreak{}CMB+\allowbreak{}gal data. The improvement in the
growth plane of the left panel of the figure is small while those in
the expansion planes of the middle and right panels are significant
due to the degeneracy breaking power of the additional data in these
planes. For this model, the combined constraints on $\Omega_{\rm
  m}=0.278^{+0.012}_{-0.011}$ and $H_0=70.0\pm1.3\Hunit$ are again
very tight.

\section{Discussion}
\label{sec:discussion}

\subsection{Comparing results}
\label{sec:comp}

For a $\Lambda$CDM expansion model, and combining galaxy and CMB data,
recent studies have presented constraints on $\gamma$ that are similar
to and in agreement with ours. For example, \cite{Hudson:12} combined
data from two peculiar velocity surveys at low redshifts
\citep{Davis:11, Turnbull:12} and RSD (but not AP effect) data from
various galaxy surveys. \cite{Samushia:12} used primarily RSD, AP
effect, and BAO data from the CMASS BOSS results of \cite{Reid:12}
together with RSD and AP effect data from other surveys. In their
combined results, both studies include WiggleZ, 6dFGS and CMASS BOSS
data, as we do here, in addition to other galaxy and expansion data
sets. Both analyses use CMB data from {\it WMAP}7. The former study
uses previous results from CMB and expansion data only as a prior,
while the latter uses the full CMB likelihood\footnote{Note that it is
  important to include $\gamma$ in the full CMB analysis, in
  combination with the other experiments, to account for all the
  degeneracies of the CMB parameters with both expansion and growth
  parameters, such as e.g. that of $\gamma$ with $\Omega_{\rm c}h^2$
  or the CMB shift parameter. If these covariances are not included,
  one may obtain spuriously tight results.}. Neither of these
analyses, however, use the low multipoles of the CMB to constrain
$\gamma$ with the ISW effect (see Section~\ref{sec:compisw}).

Note also that these studies include BAO constraints from
\cite{Percival:09} and \cite{Reid:12}, respectively. As discussed in
Section~\ref{sec:addbao}, both BAO data sets (and especially the
latter) prefer larger values for $\Omega_{\rm m}$, which in
combination with growth data implies a preference for larger values of
$\gamma$. This, together with the fact that these works do not include
the cluster data or the ISW effect constraints from the CMB data,
which both prefer smaller values for $\gamma$, is consistent with
their results on $\gamma$ being at the high end of ours in
Table~\ref{table:params}. Although all these results and those in
Table~\ref{table:params} are consistent with GR ($\gamma\simeq 0.55$),
the differences highlight the importance of studying each individual
data set as well as their various combinations in detail before
combining all of them. For upcoming, more statistically powerful data
sets, this will also be increasingly important.

\vspace{-0.1in}

\subsubsection{ISW effect}
\label{sec:compisw}

Even though the ISW effect has only a relatively small impact on the
combined results, it is not negligible. Using our analysis, it is
interesting to compare results including or not the ISW effect for
$\gamma$. For the $\gamma$+$\Lambda$CDM model and the combination
CMB+gal, we obtain $\gamma=0.607^{+0.078}_{-0.080}$ without the ISW
effect, which as expected (see Section~\ref{sec:power}) is slightly
higher ($3$ per cent) than our default result (see
Table~\ref{table:params}) and weaker by 8 per cent.\footnote{ For the
  same data but for the $\gamma$+$w$CDM model, we obtain
  $\gamma=0.547^{+0.092}_{-0.093}$, higher by $4$ per cent and weaker
  by 5 per cent, and $w=-0.914\pm0.073$, tighter by 10 per cent. For
  this model, the background expansion parameter $w$ modifies the ISW
  effect in an approximately opposite way to that of the density
  perturbation parameter $\gamma$. Therefore, by artificially ignoring
  $\gamma$ in the calculation of the ISW effect, tighter constraints
  on $w$ are obtained.  This is comparable to what happens for a
  $w$-model of a dark energy fluid when its dark energy perturbations
  are erroneously not taken into account \citep{Weller:03, Bean:03,
    Rapetti:05}.} For cl+CMB, we obtain
$\gamma=0.432^{+0.152}_{-0.153}$ and $\sigma_8=0.842\pm0.057$, which
are 20 and 16 per cent weaker than our default results \cite[see also
a similar comparison in][]{Rapetti:09}. For cl+CMB+gal,
$\gamma=0.585\pm0.067$ is only 6 per cent weaker than the
corresponding result including the ISW effect for $\gamma$.

\begin{figure*}
\begin{center}
\includegraphics[width=3.25in]{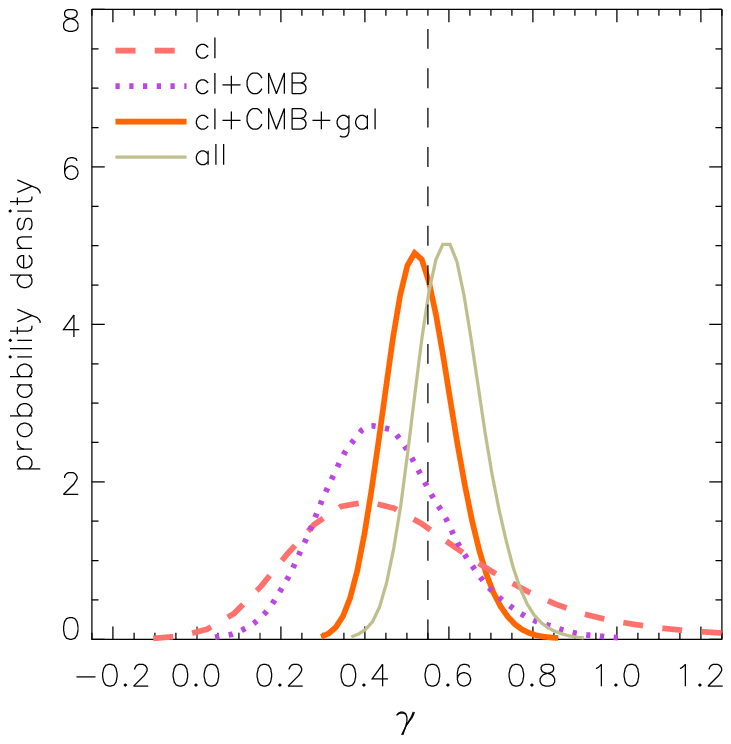}
\includegraphics[width=3.25in]{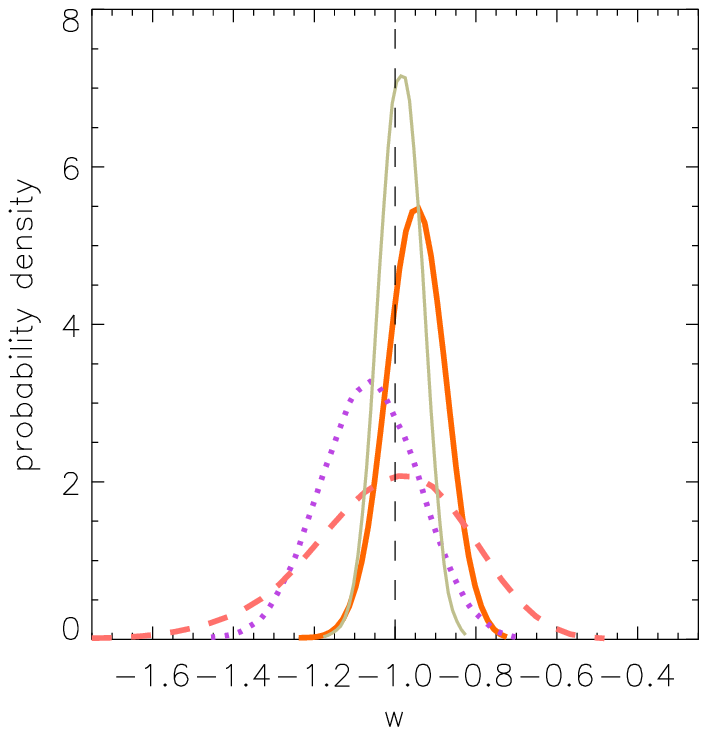}
\caption{Marginalized probability distribution functions for $\gamma$
  (left panel) and $w$ (right panel) for the $\gamma$+$w$CDM
  model. Results are shown for the following data sets: cl (red,
  dashed line), cl+CMB (purple, dotted line), cl+CMB+gal (gold, solid
  line) and cl+CMB+gal+SNIa+SH0ES+BAO (platinum, solid-thin line). The
  vertical, dashed lines mark $\gamma=0.55$ (GR; left panel) and
  $w=-1$ ($\Lambda$CDM; right panel). Note that $\gamma$ (cosmic
  growth) and $w$ (cosmic expansion) are measured simultaneously with
  similar precision.}
\label{fig:1dwcdm}
\end{center}
\end{figure*}

\vspace{-0.1in}

\subsubsection{Adding the BAO data}
\label{sec:addbao}

Our results show (see both Table~\ref{table:params} and the left panel
of Figure~\ref{fig:1dlcdm}) that, compared with the cluster and CMB
data, the combination gal+BAO prefers significantly larger values of
$\Omega_{\rm m}$, and therefore of $\gamma$ due to the covariances
between $\Omega_{\rm m}$ and $\gamma$, and $\Omega_{\rm m}$ and
$\sigma_8$ (see Section~\ref{sec:results:gammalcdm}).\footnote{For
  $\gamma$+$\Lambda$CDM, using the CMB data alone we have $\Omega_{\rm
    m}=0.260\pm0.030$. For gal+BAO, using only the BAO data set from
  \cite{Percival:09}, we obtain $\Omega_{\rm m}=0.345\pm0.050$ and
  $\gamma=0.719^{+0.244}_{-0.245}$, and using instead only the BOSS
  BAO data set, $\Omega_{\rm m}=0.417\pm0.073$ and
  $\gamma=0.978^{+0.350}_{-0.363}$, which are clearly larger.} Also,
for any of the data set combinations in Table~\ref{table:params}, the
addition of the BAO data shifts the constraints on $\Omega_{\rm m}$
and $\gamma$ to larger values. Adding BAO to all the other data sets
combined and for the $\gamma$+$w$CDM model, we have increases of $9$
and $12$ per cent for each parameter\footnote{Note that the shift
  between the gold, solid and platinum, solid-thin lines in
  Figure~\ref{fig:1dwcdm} is mainly due to the addition of BAO.}. It
is interesting to note, though, that using only the BOSS BAO data set,
we obtain similar shifts of $7$ and $11$ per cent, although slightly
weaker constraints on $\Omega_{\rm m}$, and similar constraints on
$\gamma$. Using instead only the BAO data set of \cite{Percival:09},
we find about half of those increases, $5$ per cent for both
parameters, and also a bit weaker constraints on $\Omega_{\rm m}$. The
constraints on $\gamma$, though, are slightly tighter due to the
reduction in the tension with the other data sets. The mild tension on
$\Omega_{\rm m}$ between the BAO and the other data sets translates in
some cases into a smaller constraining power for $\gamma$ (and also
for $w$) when combining them. Table~\ref{table:params} shows e.g. that
for $\gamma$+$w$CDM, adding both BAO data sets to cl+CMB+gal provides
somewhat weaker constraints on $\gamma$, and also that these are $13$
per cent weaker than those for instead adding SNIa to cl+CMB+gal. In
addition, using all the data sets combined except BAO, we obtain the
tightest constraints on $\gamma$ for $\gamma$+$\Lambda$CDM, and on
both $\gamma$ and $w$ for $\gamma$+$w$CDM. However, the increase in
constraining power on these parameters is small compared with the
decrease in constraining power on the other expansion parameters when
not using BAO.

The BAO and SH0ES data also present a mild tension in the direction of
the well-known degeneracy between $H_0$ and $\Omega_{\rm m}$ \cite[see
e.g.][]{Hinshaw:12}. The addition of SH0ES to any of the data
combinations in Table~\ref{table:params} that include our primary data
sets, shifts $H_0$ to larger values, and therefore $\Omega_{\rm m}$ to
smaller values through the correlation between these two parameters.

\subsection{Constraining power}
\label{sec:power}

As discussed in Section~\ref{sec:results:gammaw}, the combination
CMB+gal provides tight constraints on the $\sigma_8, \gamma$ plane
(see Figure~\ref{fig:datasets}) due to the complementarity between the
constraints from the individual data sets. However, the large
degeneracies of the individual constraints make the combination prone
to potential biases from systematic uncertainties. The left panel of
Figure~\ref{fig:1dlcdm} shows that for galaxies alone (green,
long-dashed line) the marginalized pdf for $\gamma$ has a large tail
toward values higher than that for GR, although interestingly the peak
is close to the GR value (vertical, dashed line).\footnote{Adding the
  expansion data set BAO to gal shortens this tail (see
  Figure~\ref{fig:datasets}) and shifts the peak to a larger value
  (see Table~\ref{table:params}).} On the other hand, for the CMB
(blue, dotted line) values larger than that for GR are significantly
constrained by the data (due to the ISW effect), while lower values
are largely unconstrained and degenerate with $\sigma_8$, which has an
extended tail toward large values (see the right panel of the figure).

From comparing the normalized pdf's in the figure, it is worth noting
that while the constraining power of the cluster data on $\gamma$
(left panel) and $\sigma_8$ (right panel) is notably better than that
of the CMB or galaxy data, the combination cl+CMB+gal is much more
powerful than the cluster data alone. Note also that the power of the
current data for constraining $\sigma_8$ (right panel) is considerably
greater than for constraining $\gamma$ (left panel).

\subsubsection{Full model: $\gamma$+$w$CDM}

For our most general model, $\gamma$+$w$CDM, only the cluster data can
alone constrain this model at a significant level. We obtain
$w=-1.021^{+0.190}_{-0.187}$ and $\gamma=0.507^{+0.236}_{-0.242}$ (see
also Figure~\ref{fig:1dwcdm}).\footnote{For the combination gal+BAO
  (see Table~\ref{table:params}), we obtain similar constraints on $w$
  and $\gamma$, while those on $\Omega_{\rm m}$ and $\sigma_8$ are
  notably weaker.} Since our other primary data sets do not have
strong direct constraints on $\gamma$ (see Section~\ref{sec:physics}),
their constraining power depends critically on the complexity of the
model used. For our extended model, we allow departures from the
standard expansion and growth histories equally. Combining all our
data sets, we obtain the tightest and most robust results to date on
this model. The addition of SNIa, SH0ES and BAO data is particularly
helpful for constraining the expansion parameters in this model. The
right panel of Figure~\ref{fig:1dwcdm} shows that when we include
these data sets (platinum, solid-thin line) the constraining power on
$w$ clearly increases. The figure also shows the progression in the
pdf's of $\gamma$ (left panel) and $w$ (right panel) when adding one
at a time the other primary data sets to the cluster data. Remarkably,
for these combinations (as well as for the others of the primary data
sets) we can measure at the same time $\gamma$ (cosmic growth) and $w$
(cosmic expansion) with similar precision.

\section{Conclusions}
\label{sec:conclusions}

We have combined cluster growth and $f_{\rm gas}$ data from RASS and
CXO, CMB data from {\it WMAP}, and RSD and AP effect data from
WiggleZ, 6dFGS and CMASS BOSS to simultaneously constrain the
evolution of cosmic structure and background expansion. To test for
consistency with GR and $\Lambda$CDM, we have used convenient
parameterizations: $\Omega_{\rm m}$, $H_0$ and $w$ for the expansion
history, and $\sigma_8$ and $\gamma$ for the growth history. We find
that the combination of clusters+CMB+galaxies breaks key degeneracies
in the growth plane, $\sigma_8$ versus $\gamma$, for the data sets
individually. In combination, the data provide tight, robust
constraints that are in excellent agreement with GR+$\Lambda$CDM.

Fixing $w=-1$, we obtain marginalized constraints on the growth
parameters $\sigma_8=0.785\pm0.019$ and
$\gamma=0.570^{+0.064}_{-0.063}$. Including SNIa, SH0ES and BAO data
we obtain $\gamma=0.616\pm0.061$. Allowing $w$ to vary, we have
$\sigma_8=0.780\pm0.020$ and $\gamma=0.533\pm0.080$ for the
combination of clusters+CMB+galaxies. For this, we find a correlation
between $w$ and $\gamma$ of $\rho=-0.66$. Including SNIa+SH0ES+BAO, we
obtain $\Omega_{\rm m}=0.278^{+0.012}_{-0.011}$,
$H_0=70.0\pm1.3\Hunit$ and $w=-0.987^{+0.054}_{-0.053}$ for the
expansion parameters, and $\sigma_8=0.789\pm0.019$ and
$\gamma=0.604\pm0.078$ for the growth parameters.

Our results highlight the potential of combining forthcoming galaxy
cluster data (from e.g. the South Pole Telescope [SPT], the Atacama
Cosmology Telescope [ACT], XMM-{\it Newton} wide area surveys, the
Dark Energy Survey [DES], the Large Synoptic Survey Telescope [LSST],
and the extended {\it ROentgen} Survey with an Imaging Telescope Array
[eROSITA]), CMB data (from e.g. SPT, ACT and {\it Planck}), and galaxy
data (from e.g. SDSS-III, the Subaru Measurement of Images and
Redshifts [SuMIRe] project, BigBOSS, the Dark Energy Spectrometer
[DESpec], and the Euclid mission) for constraining dark energy and
modified gravity models.

\section*{Acknowledgments}

We thank the anonymous referee for useful comments and G.~Morris for
technical support. The computational analysis was carried out using
the KIPAC XOC, orange and pinto clusters at SLAC. The Dark Cosmology
Centre (DARK) is funded by the Danish National Research Foundation. DR
acknowledges support from the DARK Fellowship program. CB acknowledges
the support of the Australian Research Council through the award of a
Future Fellowship. AM acknowledges support from grant NSF
AST-0838187. FB acknowledges support from the Australian Government
through the International Postgraduate Research Scholarship (IPRS). We
acknowledge support from the National Aeronautics and Space
Administration (NASA) through {\it Chandra} Award Number TM1-12010X
issued by the {\it Chandra} X-ray Observatory Center, which is
operated by the Smithsonian Astrophysical Observatory for and on
behalf of NASA under contract NAS8-03060. This work was supported in
part by the U.S. Department of Energy under contract number
DE-AC02-76SF00515. We also thank DARK and the Niels Bohr International
Academy for hospitality during the 2011 Summer workshop in which this
work was initiated.

\bibliographystyle{mnras}
\bibliography{vacpt}

\label{lastpage}
\end{document}